\documentclass[a4paper,fleqn,usenatbib]{mnras}

\usepackage{newtxtext,newtxmath}

\usepackage[T1]{fontenc}
\usepackage{ae,aecompl}

\usepackage{graphicx}	
\usepackage{amsmath}	
\usepackage{amssymb}	

\title[New \hbox{X-ray} Transient Population]{A New, Faint Population of \hbox{X-ray} Transients}

\author[F. E. Bauer et al.]{Franz E. Bauer,$^{1,2,3,4}$\thanks{E-mail: fbauer@astro.puc.cl (FEB)}
Ezequiel Treister,$^{1,4,5}$
Kevin Schawinski,$^{6}$
Steve Schulze,$^{2,1}$
\newauthor
Bin Luo,$^{7,8}$
David M. Alexander,$^{9}$
William N. Brandt,$^{10,11,12}$
Andrea Comastri,$^{13}$
\newauthor
Francisco Forster,$^{14,2}$
Roberto Gilli,$^{13}$
David Alexander Kann,$^{15}$
Keiichi Maeda,$^{16,17}$
\newauthor
Ken'ichi Nomoto,$^{17,18}$
Maurizio Paolillo,$^{19,20,21}$
Piero Ranalli,$^{22}$
\newauthor
Donald P. Schneider,$^{10,11}$
Ohad Shemmer,$^{23}$
Masaomi Tanaka,$^{24}$
Alexey Tolstov,$^{17}$
\newauthor
Nozomu Tominaga,$^{25}$
Paolo Tozzi,$^{26}$
Cristian Vignali,$^{27,13}$
Junxian Wang,$^{28}$
\newauthor
Yongquan Xue$^{28}$
and Guang Yang$^{10,11}$
\\
\\
$^{1}$Instituto de Astrof\'{\i}sica, Facultad de F\'{i}sica, Pontificia Universidad Cat\'{o}lica de Chile, 306, Santiago 22, Chile\\
$^{2}$Millennium Institute of Astrophysics (MAS), Nuncio Monse\~{n}or S\'{o}tero Sanz 100, Providencia, Santiago, Chile\\
$^{3}$Space Science Institute, 4750 Walnut Street, Suite 205, Boulder, Colorado 80301\\
$^{4}$EMBIGGEN Anillo, Concepci\'{o}n, Chile\\
$^{5}$Departamento de Astronom\'{\i}a Universidad de Concepci\'{o}n, Casilla 160-C, Concepci\'{o}n, Chile\\
$^{6}$Institute for Astronomy, Department of Physics, ETH Zurich, Wolfgang-Pauli-Strasse 27, CH-8093 Zurich, Switzerland\\
$^{7}$School of Astronomy and Space Science, Nanjing University, Nanjing 210093, China\\
$^{8}$Key laboratory of Modern Astronomy and Astrophysics (Nanjing University), Ministry of Education, Nanjing 210093, China\\
$^{9}$Department of Physics, Durham University, Durham DH1 3LE, UK\\
$^{10}$Department of Astronomy and Astrophysics, The Pennsylvania State University, 525 Davey Lab, University Park, PA 16802, USA\\
$^{11}$Institute for Gravitation and the Cosmos, The Pennsylvania State University, University Park, PA 16802, USA\\
$^{12}$Department of Physics, 104 Davey Laboratory, Pennsylvania State University, University Park, PA 16802, USA\\
$^{13}$INAF-Osservatorio Astronomico di Bologna, via Ranzani 1, 40127 Bologna, Italy\\
$^{14}$Centro de Modelamiento Matem\'{a}tico, Universidad de Chile, Av. Blanco Encalada 2120 Piso 7, Santiago, Chile\\
$^{15}$Th\"{u}ringer Landessternwarte Tautenburg, Sternwarte 5, 07778 Tautenburg, Germany\\
$^{16}$Department of Astronomy, Kyoto University, Kitashirakawa-Oiwake-cho, Sakyo-ku, Kyoto 606-8502, Japan\\
$^{17}$Kavli Institute for the Physics and Mathematics of the Universe (WPI), The University of Tokyo, Kashiwa, Chiba 277-8583, Japan\\
$^{18}$Hamamatsu Professor\\
$^{19}$Department of Physics, University Federico II, via Cintia 9, 80126, Naples, Italy\\
$^{20}$INFN - Sezione di Napoli, via Cinthia 9, 80126, Naples, Italy\\
$^{21}$Agenzia Spaziale Italiana Science Data Center, via del Politecnico snc, 00133, Roma, Italy\\
$^{22}$Lund Observatory, Department of Astronomy and Theoretical Physics, Lund University, Box 43, 22100 Lund, Sweden\\
$^{23}$Department of Physics, University of North Texas, Denton, TX 76203\\
$^{24}$National Astronomical Observatory of Japan, Mitaka, Tokyo 181-8588, Japan\\
$^{25}$Department of Physics, Faculty of Science and Engineering, Konan University, 8-9-1 Okamoto, Kobe, Hyogo 658-8501, Japan\\
$^{26}$INAF Osservatorio Astrofisico di Arcetri, Largo E. Fermi 5, IFirenze, Italy\\
$^{27}$Dipartimento di Fisica e Astronomia, Alma Mater Studiorum, Universit\'{a} degli Studi di Bologna, Viale Berti Pichat 6/2, 40127 \\ Bologna, Italy\\
$^{28}$CAS Key Laboratory for Researches in Galaxies and Cosmology, Center for Astrophysics, Department of Astronomy, University of \\ Science and Technology of China, Chinese Academy of Sciences, Hefei, Anhui 230026, China\\
\\
\\
\\
\\
\\
\\
\\
\\
\\
\\
\\
\\
\\
}

\date{Accepted February 13, 2017. Received November 8, 2016}

\pubyear{2017}

\begin{document}
\label{firstpage}
\pagerange{\pageref{firstpage}--\pageref{lastpage}}
\maketitle

\begin{abstract}
We report on the detection of a remarkable new fast high-energy
transient found in the Chandra Deep Field-South, robustly associated
with a faint (\hbox{$m_{\rm R}=27.5$\,mag}, \hbox{$z_{\rm
    ph}$$\sim$2.2}) host in the CANDELS survey. The \hbox{X-ray} event
is comprised of 115$^{+12}_{-11}$ net 0.3--7.0 keV counts, with a
light curve characterised by a $\approx$100\,s rise time, a peak
0.3--10\,keV flux of
$\approx$5$\times$10$^{-12}$\,erg\,s$^{-1}$\,cm$^{-2}$, and a
power-law decay time slope of $-1.53\pm0.27$. The average spectral
slope is $\Gamma=1.43^{+0.23}_{-0.13}$, with no clear spectral
variations. The \hbox{X-ray} and multi-wavelength properties
effectively rule out the vast majority of previously observed
high-energy transients.  A few theoretical possibilities remain: an
``orphan'' \hbox{X-ray} afterglow from an off-axis short-duration
\hbox{Gamma-ray} Burst (GRB) with weak optical emission; a
low-luminosity GRB at high redshift with no prompt emission below
$\sim$20\,keV rest-frame; or a highly beamed Tidal Disruption Event
(TDE) involving an intermediate-mass black hole and a white dwarf with
little variability. However, none of the above scenarios can
completely explain all observed properties. Although large
uncertainties exist, the implied rate of such events is comparable to
those of orphan and low-luminosity GRBs as well as rare TDEs,
implying the discovery of an untapped regime for a known transient
class, or a new type of variable phenomena whose nature remains to be
determined.
\end{abstract}




\section{Introduction}

The ever-improving depth and sky coverage of modern telescopes have
opened the floodgates to the transient universe and enabled the
discovery and characterization of several new classes of exotic
variable phenomena over the past decades
\citep[e.g.,][]{Klebesadel1973a, Bade1996a, Galama1998a,
  Kouveliotou1998a, Gezari2006a, Lorimer2007a, Soderberg2008a,
  Bloom2011a, Thornton2013a, Garnavich2016a}.
A distinct subset of such variable and transient objects can only be
understood from their high-energy properties as determined by past and
current space missions (e.g., {\it CGRO}, {\it Einstein}, {\it ROSAT},
{\it ASCA}, {\it BeppoSAX}, {\it HETE-2}, {\it RXTE}, {\it INTEGRAL},
{\it Chandra}, {\it XMM-Newton}, {\it Swift}, {\it NuSTAR}). While the
bulk of \hbox{X-ray} transients relate to accretion processes onto black
holes (BHs), neutron stars (NSs), and white dwarfs (WDs), there are
several emerging classes of exotic \hbox{X-ray} transients whose nature and
driving mechanisms remain unclear or unknown
\citep[e.g.,][]{Metzger2011a, Woosley2012a, Loeb2014a,
  Tchekhovskoy2014a, Zhang2014a, Ciolfi2016a, Irwin2016a}. Such objects provide
critical challenges to our conventional paradigms, and offer the
potential for insight into poorly understood physics.

Here we report the discovery of a new fast \hbox{X-ray} transient
found in the {\it Chandra} Deep Field-South (CDF-S). The unique
multi-wavelength properties of this transient appear to set it apart
from any known class of variable observed to date, suggesting that the
event either represents a new class of \hbox{X-ray} transient or
probes a new regime for a previously known class. While the estimated
rate of such transients remains modest and subject to large
uncertainties, their origin could have implications for future
high-energy and/or gravitational wave (GW) searches.

We have organised the paper as follows: data and analysis methods are
detailed in $\S$\ref{sec:data}; possible intrepretations are discussed
in $\S$\ref{sec:interp}; rate estimates are assessed in
$\S$\ref{sec:rates}; and finally a summary and exploration of future
prospects in $\S$\ref{sec:conclude}.  We adopt a Galactic neutral
column density of \hbox{$N_{\rm H}$$=$8.8$\times$10$^{19}$~cm$^{-2}$}
\citep{Kalberla2005a} toward the direction of the transient. Unless
stated otherwise, errors are quoted at 1$\sigma$ confidence, assuming
one parameter of interest. All magnitudes are reported in the AB
system.

\section{Data and Analysis}\label{sec:data}

We describe below the primary datasets used to detect and characterize
the transient, as well as detail the variety of constraints
obtained.

\subsection{{\it{Chandra}} 0.3--10 keV on 2014 October 01}\label{sec:chandra_xt1}

The CDF-S is the deepest survey of the \hbox{X-ray} sky, with
published observations spanning 4~Ms \citep[$\approx$46
  days;][]{Xue2011a} and an additional 3~Ms added in 2014--2016
(Chandra proposal number: 15900132; PI: W. N. Brandt). While analyzing
the new CDF-S ACIS-I \hbox{X-ray} data in near real-time, we
discovered a fast \hbox{X-ray} transient \citep{Luo2014a} midway
through one of the observations starting at 2014 October 01 07:04:37
UT (obsid 16454, $\sim$50 ks exposure). \hbox{X-ray} analysis was
performed using CIAO (v4.6) tools and custom software. Details
regarding the data processing, cleaning, photometry and alignment to
the alignment to the VLA radio and TENIS near-infrared astrometric
reference frame can be found in \citet{Xue2011a} and
\citet{Luo2017a}. We rule out all previously known {\it{Chandra}}
instrumental effects; the transient has a normal event grade and
energy distribution, and is detected in many dozens of individual
pixels tracing out portions of {\it{Chandra}}'s 32$\times$32 pixel
(16$''$$\times$16$''$) Lissajous dither pattern over a long time
duration (indicating the source is celestial).

No \hbox{X-ray}s above the background rate are detected at this position in
any other individual {\it Chandra} or {\it XMM-Newton} obsid or
combined event lists, which total 6.7\,Ms for {\it Chandra}
\citep{Luo2017a} and 2.6\,Ms for {\it XMM-Newton} \citep{Comastri2011a}.
Here we assumed the 90\% encircled energy region to derive the source
limit, while for {\it XMM-Newton} we adopted a circular aperture of
6$''$ radius.  These limits imply quiescent 0.3--10, 0.3--2.0, and
2--10\,keV flux limits of
3.1$\times$10$^{-17}$,
1.6$\times$10$^{-17}$, and
5.4$\times$10$^{-17}$
erg\,cm$^{-2}$\,s$^{-1}$, respectively, at 3$\sigma$ confidence.  The
first count from \hbox{CDF-S XT1} arrives $\approx$16.8\,ks into the
observation, offering immediate precursor 0.3--10, 0.3--2.0, and
2--10\,keV flux limits of
5.2$\times$10$^{-15}$,
2.9$\times$10$^{-15}$, and
7.9$\times$10$^{-15}$
erg\,cm$^{-2}$\,s$^{-1}$, respectively.

\begin{figure}
\vspace{0in}
\begin{center}
\hglue-0.2cm{\includegraphics[width=3.8in]{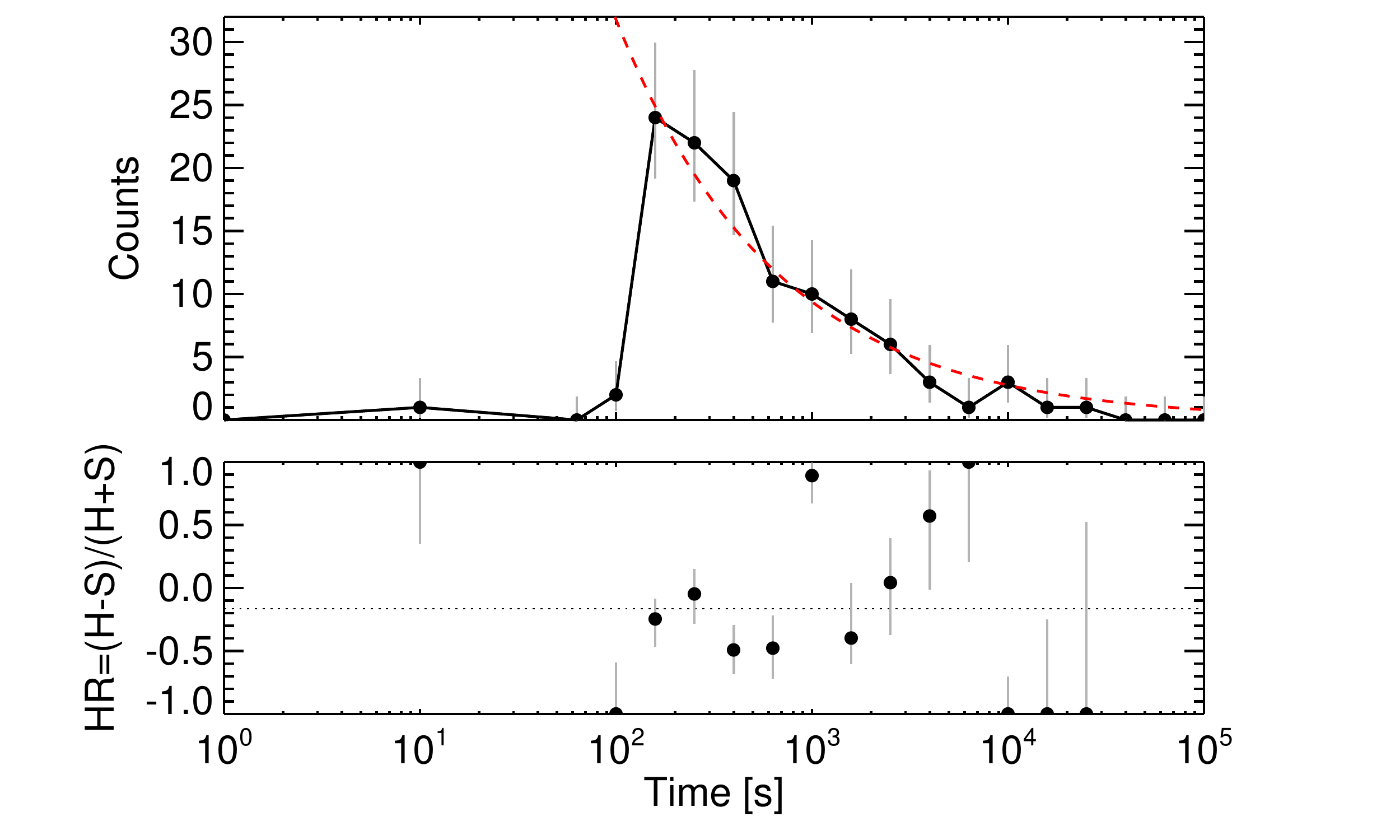}}
\end{center}
\vspace{0in}
\caption{\hbox{X-ray} light curve (top panel) and hardness ratio
  (bottom panel) of \hbox{CDF-S XT1}. To highlight the sharp rise at
  $\approx$110\,s, the 0.3--7.0 keV counts are logarithmically binned
  and shown with 1$\sigma$ errors \citep{Gehrels1986a}; for this
  reason, binning here differs somewhat from that provided in
  Table~\ref{tab:xray_ltcrv}. The red dashed curve denotes the
  best-fitting powerlaw decay time slope of $a$$=$$-1.53$. The
  hardness ratio, HR, and 1$\sigma$ errors are calculated as
  ($H-S$)/($H+S$) following the Bayesian method of \citep{Park2006a},
  where S and H correspond to the 0.3--2.0\,keV and 2.0--7.0\,keV
  counts, respectively. We omit bins with no counts in the bottom
  panel, since HR values are completely unconstrained. The dotted
  horizontal line signifies the HR value expected for a
  $\Gamma$$=$1.43 power law.}
\vspace{-0.1in}
\label{fig:xraylc}
\end{figure}

\begin{figure*}
\vspace{0in}
\begin{center}
\hglue-0.2cm{\includegraphics[width=3.5in]{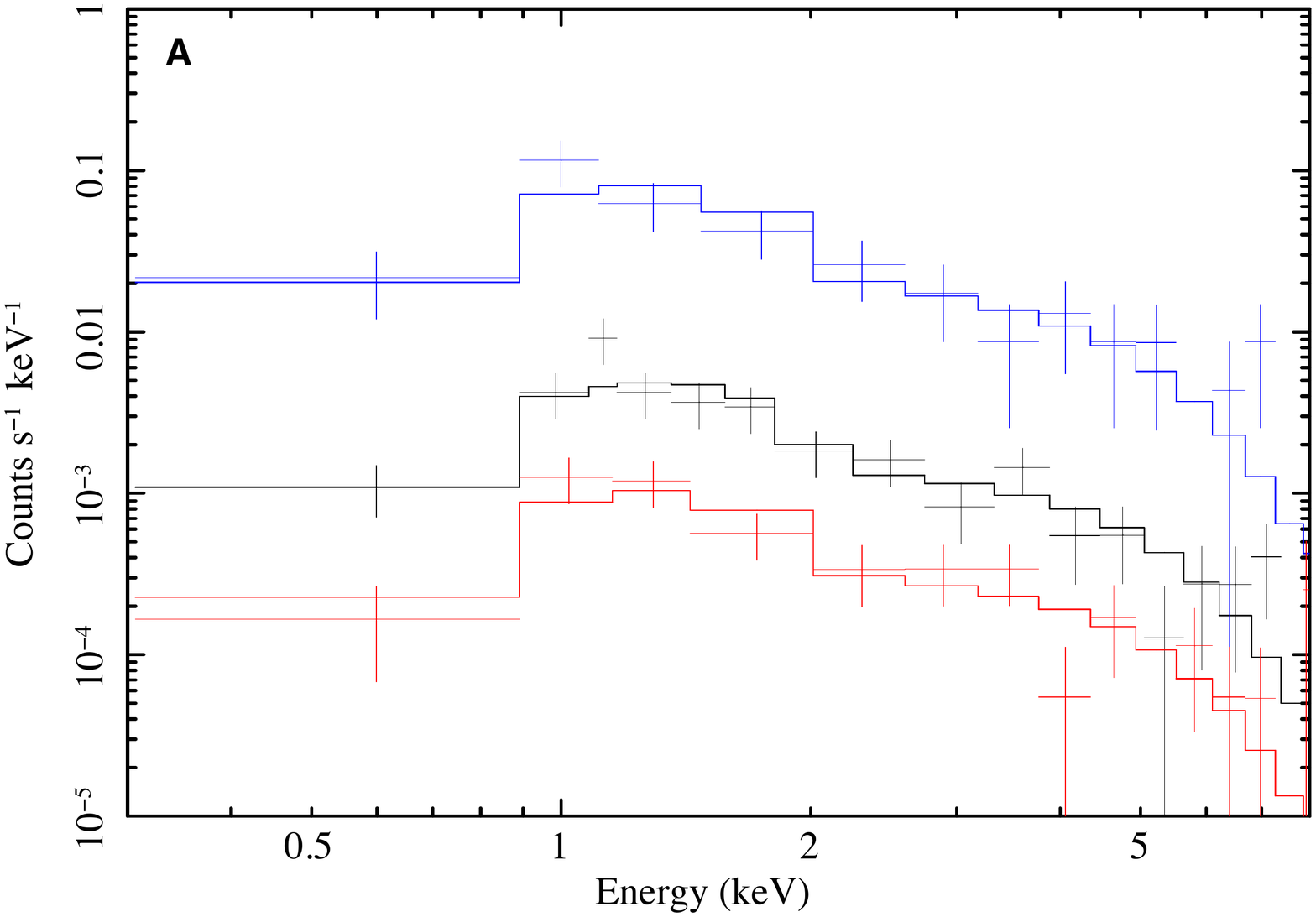}}\hfill
\hglue-1.0cm{\includegraphics[width=3.5in]{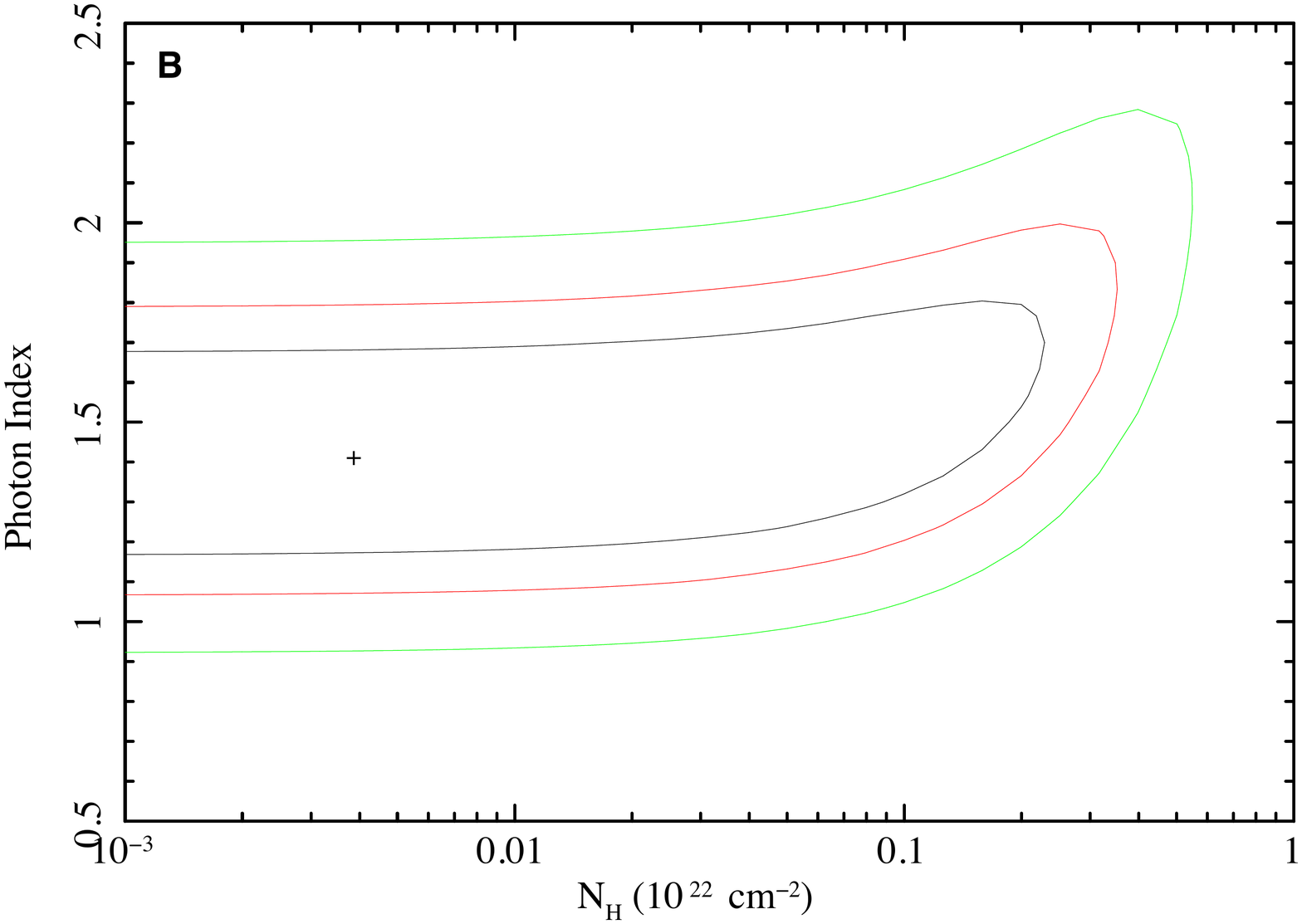}}
\end{center}
\vspace{0in}
\caption{Best-fitting \hbox{X-ray} spectral model for \hbox{CDF-S
    XT1}. {\bf (A)} \hbox{X-ray} spectra (data points) and
  best-fitting power-law models (histogram), where the complete
  spectrum is denoted in black, while the early ($<$400\,s) and late
  ($>$400\,s) spectra are shown in blue and red, respectively. The
  best-fitting power-law model to the complete spectrum yields
  $\Gamma$$=$1.43$^{+0.26}_{-0.15}$ and $N_{\rm
    H}$$<$1.5$\times$10$^{21}$ $(1+z)^{2.5}$ cm$^{-2}$. No significant
  evidence for spectral hardening is seen between the two epochs. {\bf
    (B)} Confidence contours of $\Gamma$ and $N_{\rm H}$ when modeled
  for the complete spectrum (1$\sigma$ black, 2$\sigma$ red, 3$\sigma$
  green).}
\vspace{-0.1in}
\label{fig:xrayspec}
\end{figure*}

The transient has a J2000 position of
\hbox{$\alpha$=53\fdg161550},
\hbox{$\delta$=-27\fdg859353} and an estimated 1$\sigma$
positional uncertainty of 0\farcs26. We extracted 115 net
counts in the 0.3--7 keV band for the transient from a 3$''$ radius
circular aperture (97\% encircled energy fraction at 1.5 keV at the
source position and zero expected background counts), from which we
constructed an \hbox{X-ray} light curve (Fig.~\ref{fig:xraylc}) and spectrum
(Fig.~\ref{fig:xrayspec}) following standard procedures. We
arbitrarily set the light curve zeropoint as 10\,s prior to the
arrival of the first photon.

The count rate of the transient near the peak of its \hbox{X-ray} light curve
is $\approx$0.3--0.4 cts s$^{-1}$ (equivalent to a readout of
$\gtrsim$1\,count
per 3.2\,s frame). As such, there is some potential for photons to
suffer pile-up (two incident photons count as one higher energy photon
or possibly even rejected), which would harden the spectrum and lower
the observed count rate at early times. Fortunately, the transient
lies at an off-axis angle of 4\farcm3 from the ACIS-I
aimpoint, and thus has an extended point spread function (PSF), such
that only $\approx$50\% of the photons lie within
$\approx$1\farcs0--1\farcs2 (for energies of
1.5--6.4 keV, respectively). Such a PSF should be $\approx$4 times
less affected by pileup compared to an on-axis PSF, implying that
\hbox{CDF-S XT1} should be minimally affected by pileup (a few percent
at most). We verified this result empirically by examining the source
frame by frame and based on simulations with the MARX
\citep{Davis2012a}.

The \hbox{X-ray} light curve (Fig.~\ref{fig:xraylc}) shows a fast rise
of 110$\pm$50\,s to a peak 0.3--10 keV flux of
$5.1\times10^{-12}$\,erg\,s$^{-1}$\,cm$^{-2}$ and a power-law decay of
the form $F_{\rm 0.3-10 keV}=F_{0} (t/t_{0})^{a}$, where $t$ is time,
$F_{0}$$=$$0.364\pm0.083$\,erg\,s$^{-1}$\,cm$^{-2}$,
$t_{0}$$=$$146\pm12$\,s, and slope $a$$=$$-1.53\pm0.27$, fit using a
least-squares method. Dividing the light curve into logarithmic time
bins of 0.2\,dex, we see marginal evidence for spectral hardening in
one bin around 1000\,s. However this is not strong enough to rule out
a constant model at $>$3$\sigma$, and thus no significant spectral
variations with time are found given the limited statistics
(Fig.~\ref{fig:xraylc}; see also Table~\ref{tab:xray_ltcrv} using
Bayesian block binning). The $T_{\rm 90}$ duration parameter, which
measures the time over which the event emits from 5\% to 95\% of its
total measured counts, is $5.0^{+4.2}_{-0.3}$\,ks, with an associated
fluence of ($4.2^{+3.5}_{-0.2}$)$\times$10$^{-9}$ erg\,cm$^{-2}$.

\begin{table}
\centering
\begin{tabular}{rrccc}
\hline
Time & Bin Half & CR & $F_{\rm 0.3-10\,keV}$ & HR \\
Bin &  Width & & & \\
\hline
    16 &    16 &  $4.24\times10^{-2}$ & $8.5^{+15.4}_{-6.4}\times10^{-13}$ & \llap{$-$}$0.23^{+0.74}_{-0.43}$\\
    64 &    32 &  $3.82\times10^{-2}$ & $7.7^{+ 8.8}_{-4.6}\times10^{-13}$ & \llap{$-$}$0.50^{+0.49}_{-0.43}$\\
   160 &    64 &  $2.45\times10^{-1}$ & $5.1^{+ 1.1}_{-0.9}\times10^{-12}$ & \llap{$-$}$0.21^{+0.16}_{-0.18}$\\
   352 &   128 &  $1.14\times10^{-1}$ & $2.3^{+ 0.5}_{-0.4}\times10^{-12}$ & \llap{$-$}$0.14^{+0.14}_{-0.21}$\\
   832 &   352 &  $3.59\times10^{-2}$ & $7.2^{+ 1.7}_{-1.4}\times10^{-13}$ &           $0.07^{+0.16}_{-0.24}$\\
  1664 &   480 &  $1.03\times10^{-2}$ & $2.1^{+ 0.9}_{-0.6}\times10^{-13}$ & \llap{$-$}$0.19^{+0.35}_{-0.25}$\\
  3568 &  1424 &  $3.19\times10^{-3}$ & $6.4^{+ 2.9}_{-2.1}\times10^{-14}$ &           $0.02^{+0.48}_{-0.33}$\\
 18000 & 13008 &  $2.69\times10^{-4}$ & $5.4^{+ 2.9}_{-2.0}\times10^{-15}$ & \llap{$-$}$1.00^{+1.42}_{-0.00}$\\
138320 & 12496 &  $3.70\times10^{-5}$ & $7.5^{+18.7}_{-6.9}\times10^{-16}$ & \llap{$-$}$1.00^{+1.50}_{-0.00}$\\
\hline
\end{tabular}
\caption{
\hbox{X-ray} timing properties of \hbox{CDF-S XT1}.
{\it Col. 1}: Bayesian block time bin, in seconds, following \citep{Scargle2013a}.
{\it Col. 2}: Time bin half width, in seconds.
{\it Col. 3}: 0.3--7\,keV count rate, in cnts s$^{-1}$.
{\it Col. 4}: 0.3--10\,keV flux, in erg\,cm$^{-2}$\,s$^{-1}$.
{\it Col. 5}: Hardness ratio (HR) defined as ($H-S$)/($H+S$) where $H$ and $S$
are the 2--7 and 0.3--2\,keV counts, respectively \citep{Park2006a}.
\label{tab:xray_ltcrv}}
\end{table}

Given the low number of counts, the \hbox{X-ray} spectrum was fit
using the Cash statistic \citep{Cash1979a} with relatively simple
continuum models. The data are well-fitted by either an absorbed power
law ($dN/dE\propto E^{-\Gamma}$) with
$\Gamma$$=$1.43$^{+0.26}_{-0.15}$ or a relatively unconstrained
20.2$^{+27.6}_{-11.3}$\,keV absorbed thermal plasma ({\tt apec})
model, with a best-fitting absorption limit of $N_{\rm
  H}$$<$4.5$\times$10$^{21}$\,$(1+z)^{2.5}$ cm$^{-2}$ (3$\sigma$). The
latter is formally consistent with the Galactic value of
8.8$\times$10$^{19}$\,cm$^{-2}$ \citep{Kalberla2005a}, but intrinsic
absorption cannot be ruled out. These values should be used with
caution, as there is some possible degeneracy between the best-fitting
photon index $\Gamma$ and column density $N_{\rm H}$, such that a
softer value of $\Gamma$ ($\sim$2) cannot be excluded; see
Fig.~\ref{fig:xrayspec}. The absorption limit implies $A_{\rm
  V}$$\lesssim$0.7\,$(1+z)^{2.5}$\,mag assuming a Galactic dust-to-gas
ratio \citep{Guver2009a}, which could become substantial at large
distances due to the strong redshift dependence.  The observed 0.3--10
(2--10)\,keV flux from the total spectrum, which we extracted from the
first 12\,ks only to optimise the inclusion of source versus
background photons, is $2.0\times10^{-13}$ ($1.4\times10^{-13}$)
erg\,cm$^{-2}$\,s$^{-1}$.

To investigate further the possibility of spectral variance with time,
we split the spectrum in two parts with roughly equal photon counts:
$<$400\,s (``early'') and $>$400\,s (``late''). This cut roughly
coincides with the harder tine bin at $\sim$1000\,s seen in
Fig.~\ref{fig:xraylc}. The best-fitting absorbed power law models
yielded $\Gamma_{\rm early}=1.63^{+0.42}_{-0.21}$ and $N_{\rm H,
  early}$$<$$7.2\times10^{21} (1+z)^{2.5}$ cm$^{-2}$ at early times
and $\Gamma_{\rm late}=1.50^{+0.42}_{-0.33}$ and $N_{\rm H,
  late}$$<$$1.0\times10^{22} (1+z)^{2.5}$ cm$^{-2}$ at late times,
respectively (3$\sigma$). The corresponding observed early and late
time fluxes at 0.3--10 (2--10) keV are $2.9\times10^{-12}$
($1.8\times10^{-12}$) and $4.0\times10^{-14}$ ($2.9\times10^{-14}$ )
erg\,cm$^{-2}$\,s$^{-1}$, respectively; we find a factor of
$\approx$70 decrease in the 0.3--10 keV flux between the early and
late regimes. If the column density $N_{\rm H}$ is left free but
required to be the same at early and late times, there is no change to
the early time slope while the late time spectral index drops to
$\Gamma_{\rm late}=1.41^{+0.33}_{-0.23}$ with $N_{\rm H, early+late}
<5.5\times10^{21} (1+z)^{2.5}$ cm$^{-2}$ (3$\sigma$).  Again, there is
no evidence for significant spectral hardening of the transient with
time, within the statistical limitations of the data.  Based on the
confidence contours assessed for the complete spectrum, the source is
consistent with Galactic absorption only, although it could be
absorbed by as much as $7.2\times10^{21} (1+z)^{2.5}$ cm$^{-2}$ at
early times. The 2--10 keV \hbox{X-ray} luminosity for a variety of
redshifts is provided in Table~\ref{tab:energetics}.

\subsection{Previous imaging}\label{sec:chandra_all}

The high Galactic latitude ($l$$=$223$^{\circ}$, $b$$=$$-54^{\circ}$),
low extinction CDF-S region has been the subject of many intensive
observing campaigns, and has some of the deepest coverage to date at
nearly all observable wavelengths. We used in particular the images
from the {\it Hubble Space Telescope} ({\it{HST}}) GOODS
\citep{Giavalisco2004a} and CANDELS \citep{Grogin2011a, Koekemoer2011a}
surveys to identify and constrain the potential host galaxy of the
\hbox{X-ray} transient. Both the CANDELS F160W DR1 \citep{Guo2013a} and
3D-{\it{HST}} v4.1 \citep{Skelton2014a} catalogs detect several sources
in the vicinity of the \hbox{X-ray} transient with comparable
brightnesses. We adopt values from CANDELS, which provides TFIT
\citep{Laidler2007a} photometry measured on calibrated images, while
3D-{\it{HST}} fit their spectroscopic data with a set of templates and
then correct the photometry; there are magnitude differences as large
as $\sim$1 mag between catalogs, as well as detections in CANDELS but
not in 3D-{\it{HST}}, despite clear visual
confirmation. Table~\ref{tab:optids} lists the optical sources in
the vicinity of the \hbox{X-ray} transient, in order of distance.

\begin{table*}
{\scriptsize
\centering
\begin{tabular}{lllllllllllll}
\hline
\# & CANDELS & R.A, Dec. & offset & $R$   & $J$   & $r_{\rm Kron}$ & $z_{\rm ph}$ & $M_{R}$ & $\log(M/M_{\odot})$ & SFR ($M_{\odot}$\,yr$^{-1}$) \\
\hline
1 & 28573 & 53.161575, -27.859375 & 0.13  & 27.51 & 27.31 & 0.56  & 2.23 (0.39--3.21) & -18.7  & $7.99\pm0.20$ & $1.15\pm0.04$ \\
2 & 28572 & 53.161841, -27.859427 & 1.09  & 27.38 & 27.34 & 0.50$^{a}$ & 0.31 (0.07--6.81)$^{b}$ & -13.7  & $6.84\pm0.18$ & $0.40\pm0.07$ \\
3 &  5438 & 53.161095, -27.859668 & 1.99  & 26.89 & 26.16 & 0.76  & 0.53 (0.18--2.89) & -15.5  & $7.82\pm0.33$ & $1.20\pm0.60$ \\
4 &  5448 & 53.162391, -27.859707 & 3.29  & 25.78 & 25.27 & 0.69  & 0.15 (0.10--0.26) & -13.5  & $7.47\pm0.12$ & $0.03\pm0.01$ \\
\hline
\end{tabular}
\caption{
 CANDELS F160W Data Release 1 (DR1) catalog parameters
{\it Col. 1}: Object number.
{\it Col. 2}: CANDELS catalog number. For completeness, the
equivalent 3D-{\it{HST}} v4.1 catalog numbers are 10718, 10709, 10670, and
10685, respectively.
{\it Col. 3}: J2000 Right Ascension and Declination in degrees.
{\it Col. 4}: Angular offset between \hbox{X-ray} and optical positions.
{\it Col. 5}: observed $R$-band magnitude.
{\it Col. 6}: observed $J$-band magnitude.
{\it Col. 7}: Kron radius. $^{a}$ FWHM$\approx$0\farcs12 is consistent with point source.
{\it Col. 8}: Photometry redshift and 95\% limit range in parentheses
from best-fitting template. $^{b}$ The photometry for 28572 can also
be fit with a Galactic M-star template.
{\it Col. 9}: Absolute $R$-band magnitude.
{\it Col. 10}: Estimated logarithm of stellar mass at best-fitting $z_{\rm ph}$.
{\it Col. 11}: Estimated star formation rate (SFR) at best-fitting $z_{\rm ph}$.
\label{tab:optids}}
}
\end{table*}

\begin{figure*}
\vspace{0in}
\begin{center}
\includegraphics[width=7.0in]{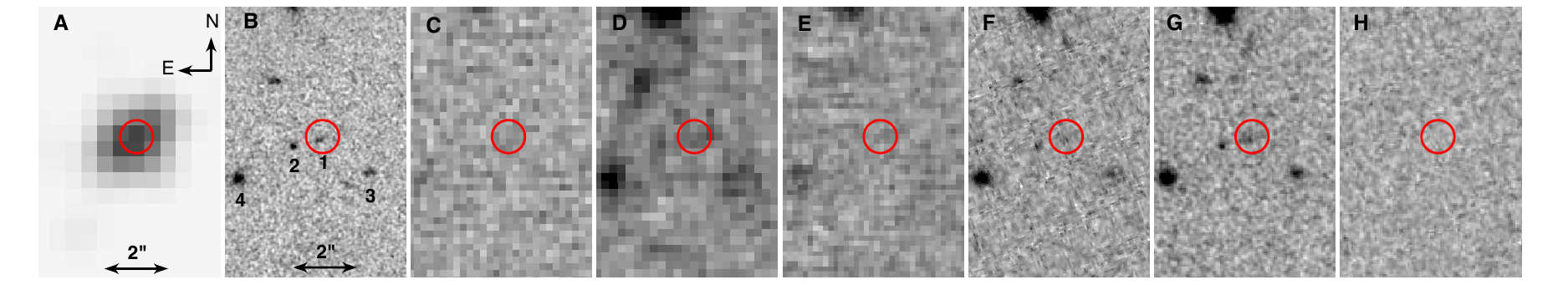}
\end{center}
\vspace{0in}
\caption{Images (6$''$$\times$11$''$) in the vicinity of \hbox{CDF-S XT1}. From left to right:
{\bf (A)} {\it{Chandra}} ACIS-I 0.3--7.0\,keV image of the transient detection acquired on 2014 October 01;
{\bf (B)} {\it{HST}}/ACS $F606W$ image from GOODS-S acquired prior to 2008 \citep{Giavalisco2004a};
{\bf (C)} VLT/VIMOS $R$-band image serendipitously acquired on 2014 October 01 (80\,min post-transient);
{\bf (D)} VLT/FORS2 $R$-band image acquired on 2014 October 18 (18 days post-transient); 
{\bf (E)} Gemini/GMOS-S $r$-band image acquired on 2014 October 29 (27 days post-transient)
{\bf (F)} {\it{HST}}/WFC3 $F110W$ image acquired on 2015 January 20 (111 days post-transient). 
{\bf (G)} {\it{HST}}/WFC3 $F125W$ image from CANDELS acquired prior to 2011 \citep{Grogin2011a, Koekemoer2011a}.
{\bf (H)} $F110W-F125W$ difference image.
A 0\farcs52 radius red circle denotes the 2$\sigma$
\hbox{X-ray} positional error, centred on the \hbox{X-ray} transient position. The
closest potential optical counterpart, seen clearly in the {\it{HST}}
images and labeled \#1 in (B), lies 0\farcs13 southeast of the
\hbox{X-ray} position and has a magnitude of $m_{\rm R}$$=$27.5\,mag. It is
classified as a dwarf galaxy with $z_{\rm ph}$$=$2.23. This
galaxy appears marginally detected in the 1\,hr FORS2 image, but not
in the VIMOS or GMOS-S images. Three other sources are labeled and
discussed in the text. 
No transient is observed in the {\it{HST}} difference image (final
panel). }
\vspace{-0.1in}
\label{fig:opt_img}
\end{figure*}

Given the error in the \hbox{X-ray} position, source \#1 is clearly
the favoured counterpart and we can exclude all other detected sources
at $\gtrsim$4$\sigma$. Based on the source density of the CANDELS
$F160W$-band catalog \citep{Guo2013a}, the probability of a random
alignment between \hbox{CDF-S XT1} and a source as bright as \#1
within a radius of 0\farcs13 is $<$0.1\%.\footnote{Even adopting a
  3$\sigma$ radius of 0\farcs78, the probability of a random match
  remains quite low ($<$4\%).}  At the best-fitting photometric
redshift of $z_{\rm ph}$$=$2.23, SED fitting of the CANDELS DR1
optical/NIR photometry suggests that the nearest counterpart is a
dwarf galaxy with $M_{R}$$=$$-17.3$\,mag (i.e., a few times smaller
than the Large Magellanic Cloud, but with a stronger star formation
rate of 1.5\,$M_{\odot}$\,yr$^{-1}$). The 1$\sigma$ and 2$\sigma$
ranges on the photometric redshift from the CANDELS $F160W$-band
catalog are 1.57--2.81 and 0.39--3.21, respectively. The reported
1$\sigma$ errors on the other derived properties listed in
Table~\ref{tab:optids} are only statistical, measured at the
best-fitting photometric redshift, which is fixed. Incorporating the
$z_{\rm ph}$ error distribution and other systematic errors, which are
difficult to quantify, are likely to increase the errors
substantially.  The absolute magnitude of the host for a variety of
redshifts is provided in Table~\ref{tab:energetics}.  The host does
not appear to be particularly dusty. Three $R$-band images and one
{\it{HST}} Wide Field Camera 3 (WFC3) $F110W$-band image of the field
have been acquired since the \hbox{X-ray} detection, spanning 0.06 and
111 days post-transient (see Fig.~\ref{fig:opt_img}). As outlined in
$\S$\ref{sec:vimos}--\ref{sec:hst}, no clear transient counterpart is
detected at $m_{\rm R}$$\lesssim$25.5--26.5\,mag \citep{Luo2014a,
  Treister2014a, Treister2014b}
in the optical and $\lesssim28.4$\,mag in the $F110W$ band.

\subsection{VLT/VIMOS $R$ on 2014 October 1 (E1)}\label{sec:vimos}

Serendipitously, the field of the fast \hbox{X-ray} transient was observed
almost simultaneously ($\approx$80 minutes after) at optical
wavelengths by the 8.2m Very Large Telescope (VLT) of the European
Southern Observatory (ESO) using the VIsible MultiObject Spectrograph
(VIMOS), as part of the VANDELS\footnote{http://vandels.inaf.it}
public survey (PIs: R. McLure and L. Pentericci). A 550\,s $R$-band
image (program ID 194.A-2003A) was obtained starting at 2014
October 1 08:20:09.6 UT with an optical seeing of $\approx$0\farcs7
FWHM and an average airmass of 1.03 (hereafter epoch 'E1'). The data
were retrieved from the ESO archive and reduced using standard
procedures. After aligning the \hbox{X-ray} and $R$-band images to
$\approx$0\farcs1, no object is detected at the location of the \hbox{X-ray}
flare, with an estimated magnitude limit of $m_{\rm R}$$\approx$25.7\,mag
(2$\sigma$, 0\farcs5 radius aperture).  There is evidence for a
marginal detection of source \#4, as seen in the deep {\it{HST}} data, but
nothing fainter. No variable sources are found within at least
20--30$''$ of the \hbox{X-ray} transient. The absolute magnitude limit of the
transient for a variety of redshifts is provided in
Table~\ref{tab:energetics}.

\subsection{VLT/FORS2 $R$ on 2014 October 19 (E2)}\label{sec:fors}

Following the {\it{Chandra}} detection and initial serendipitous
observation, the field of the fast \hbox{X-ray} transient was observed
again with the 8.2m VLT using the FOcal Reducer and low dispersion
Spectrograph (FORS2), 18 days after the \hbox{X-ray} transient was
detected, as part of DDT program 294.A-5005A (PI: Franz Bauer). A
2900\,s $R$-band image was obtained starting at 2014 October 19
05:37:15.6 UT under photometric conditions with an optical seeing of
$\approx$0\farcs8 FWHM in the optical and an average airmass of 1.01
(hereafter epoch 'E2'). The $R$ filter was chosen as a compromise
between the potential expectation for a blue transient, possible
obscuration, and the relative sensitivity of the detector. The
$\sim$7'$\times$7' field of view covered by FORS2 was centred on the
reported coordinates of the \hbox{X-ray} transient. The data were
retrieved from the ESO archive and reduced using standard
procedures. After aligning the \hbox{X-ray} and $R$-band images to
$\approx$0\farcs1, no source is formally detected at the position of
the \hbox{X-ray} transient, with an estimated magnitude limit of
$m_{\rm R}$$\approx$27.0\,mag (2$\sigma$, 0\farcs5 radius
aperture). The nearest detected source is \#3.  No variable sources
are apparent within at least 20--30$''$ of the \hbox{X-ray}
transient. The absolute magnitude limit of the transient for a variety
of redshifts is provided in Table~\ref{tab:energetics}.

\subsection{Gemini-S/GMOS-S $r$ on 2014 October 28 (E3)}\label{sec:gmos}

The field of the \hbox{X-ray} transient was observed a third time by
the 8m Gemini-South Telescope using the imager on the Gemini
Multi-Object Spectrograph (GMOS-S), 27 days after the \hbox{X-ray}
transient was detected, as part of DDT program GS-2014B-DD-4 (PI:
Ezequiel Treister). A 4500\,s $r$-band image was obtained starting at
2014 October 28 07:36:25.7 UT under clear conditions with an optical
seeing of $\approx$0\farcs6 FWHM and an average airmass of 1.2
(hereafter epoch 'E3'). The $r$ filter was chosen as a compromise
between the potential expectation for a blue transient, possible
obscuration, the relative sensitivity of the detector, and to crudely
match the previous two observations. With the new Hamamatsu CCDs
installed, GMOS-S covers a $\approx$5\farcm5$\times$5\farcm5 field of
view, which was centred on the \hbox{X-ray} transient. The data were
retrieved from the Gemini archive and reduced using standard
procedures. After aligning the \hbox{X-ray} and $r$-band images to
$\approx$0\farcs1, no source is formally detected at the position of
the \hbox{X-ray} transient, with an estimated magnitude limit of
$m_{\rm r}$$\approx$26.0\,mag (2$\sigma$, 0\farcs5 radius
aperture).\footnote{This limit is roughly 0.6 mag brighter than
  estimated by the Gemini-South GMOS-S ITC, possibly due to early
  background problems associated with the newly installed Hamamatsu
  CCDs.}  The nearest detected source to the \hbox{X-ray} position is
\#4. No variable sources are apparent within at least 20--30$''$ of
the \hbox{X-ray} transient. The absolute magnitude limit of the
transient for a variety of redshifts is provided in
Table~\ref{tab:energetics}.

\subsection{{\it{HST}}/WFC3 $F110W$ on 2015 January 20 (E4)}\label{sec:hst}

The field of the \hbox{X-ray} transient was observed a fourth time by
{\it{HST}} using WFC3, 111 days after the \hbox{X-ray} transient was
detected, as part of DDT program HST-GO-14043 (PI: Franz Bauer). A
2612\,s $F110W$-band image was obtained on 2015 January 20 11:00:13
UT, with a dithered field of view of $\approx$140$''$$\times$124$''$
centred on the \hbox{X-ray} transient (hereafter epoch 'E4'). We
switched to the $F110W$ filter to test whether the \hbox{X-ray}
transient might have been exceptionally red due to strong extinction
or high-redshift, and owing to the excellent sensitivity of this band
for faint NIR emission.  The data were retrieved from the Mikulski
Archive for Space Telescopes and reduced using standard
procedures. After aligning the \hbox{X-ray} and $F110W$-band images to
$\approx$0\farcs1, we recover nearly all of the objects from the deep
{\it{HST}} $F125W$ image of CANDELS, including the associated
counterpart source \#1, with $m_{\rm F110W}$$=$27.43\,mag. Based on
difference imaging with the CANDELS $F105W$ and $F125W$ images using
the High Order Transform of PSF and Template Subtraction code
\citep[HOTPANTS;][]{Becker2015a}, we place a limit of $m_{\rm
  F110W}$$\approx$28.4\,mag (2$\sigma$, 0\farcs2 radius aperture),
comparable to the expected magnitude limit based on the {\it HST}
exposure time calculator. To place this value in context for
Fig.~\ref{fig:optgrb}, we assume a colour dependence of \hbox{$m_{\rm
    R}$$-$$m_{\rm F110W}$$\approx$0.4--0.7\,mag} based on GRB afterglow
power-law spectral slopes in the range of $-$0.6 to $-$1.1
\citep{Kann2010a,Kann2011a} and 0.4--1.0 mag for CCSNe between
$z=0.0$--1.0 \citep{Poznanski2002a,Drout2011a,Bianco2014a}.  This
implies an equivalent limit of $m_{\rm R}$$=$28.8--29.4\,mag. Again,
no variable sources are detected within at least 20--30$''$ of the
\hbox{X-ray} transient. The absolute magnitude limit of the transient
from the difference imaging is provided in Table~\ref{tab:energetics}
for a variety of redshifts.

\begin{table*}
\centering
\begin{tabular}{lllllll}
\hline
Redshift & $L_{\rm 2-10\,keV}$ & $M_{\rm F606W, Host}$ & $M_{\rm R, E1}$ & $M_{\rm R, E2}$ & $M_{\rm r, E3}$ & $M_{\rm F110W, E4}$ \\
         & ($10^{45}$ erg\,s$^{-1}$) & (mag) & (mag) & (mag) & (mag) & (mag)\\
\hline
2.23 &  67.5  & $-$18.7 & $>$$-$20.5 & $>$$-$19.3 & $>$$-$20.3 & $>$$-$17.9\\
\hline
0.30 &   0.5  & $-$13.5 & $>$$-$15.3 & $>$$-$14.0 & $>$$-$15.0 & $>$$-$12.6 \\
0.50 &   1.7  & $-$14.8 & $>$$-$16.6 & $>$$-$15.3 & $>$$-$16.3 & $>$$-$13.9 \\
1.00 &   9.4  & $-$16.6 & $>$$-$18.4 & $>$$-$17.1 & $>$$-$18.1 & $>$$-$15.7 \\
2.00 &  51.7  & $-$18.5 & $>$$-$20.3 & $>$$-$19.0 & $>$$-$20.0 & $>$$-$17.6 \\
3.00 & 138.5  & $-$19.5 & $>$$-$21.3 & $>$$-$20.0 & $>$$-$21.0 & $>$$-$18.6 \\
\hline
\end{tabular}
\caption{
Intrinsic constraints on \hbox{CDF-S XT1} for several example redshifts.
{\it Col. 1}: Example redshift. The photometric redshift of the
nearby galaxy is nominally 2.23, but extends between 0.39--3.21 at
95\%, so we provide a wide range.
{\it Col. 2}: 2--10\,keV \hbox{X-ray} luminosity within the initial 400\,s,
in units of $10^{45}$\,erg\,s$^{-1}$.
{\it Col. 3}: Absolute $F606W$-band magnitude of the tentative host galaxy.
{\it Col. 4}: Absolute $R$-band magnitude limit for epoch E1.
{\it Col. 5}: Absolute $R$-band magnitude limit for epoch E2.
{\it Col. 6}: Absolute $r'$-band magnitude limit for epoch E3.
{\it Col. 7}: Absolute $F110W$-band magnitude limit for epoch E4.
\label{tab:energetics}}
\end{table*}

\subsection{ATCA/CABB 2--19\,GHz on 2014 October 08}\label{sec:atca}

Radio observations of the field of the \hbox{X-ray} transient were made on
2014 October 08 with the Australian Telescope Compact Array (ATCA) in
the 1.5\,km configuration using the Compact Array Broadband Backend
(CABB) at 2.1, 5, 9, 17, and 19 GHz \citep{Burlon2014a}. No radio
emission was detected in the vicinity of the transient, with $3\sigma$
limits of $S_{\rm 2.1\,GHz}<174$\,$\mu$Jy, $S_{\rm                                                                
  5\,GHz}<81$\,$\mu$Jy, $S_{\rm 9\,GHz}<75$\,$\mu$Jy, $S_{\rm                                                     
  17\,GHz}<105$\,$\mu$Jy, and $S_{\rm 19\,GHz}<99$\,$\mu$Jy,
respectively. Radio limits based on observations obtained prior to the
transient are $S_{\rm                                                                                             
  1.4\,GHz}$$\lesssim$24\,$\mu$Jy
and $S_{\rm                                                                                                       
  5\,GHz}$$\lesssim$27\,$\mu$Jy
\citep{Miller2013a, Burlon2014a}.

\section{Possible Interpretations}\label{sec:interp}

We detail below a set of possible scenarios that might produce a fast
\hbox{X-ray} transient such as the one we observed. In many cases, we are
able to exclude these scenarios based on our available
multi-wavelength constraints. This list may not account for every
possibility and should not be interpreted as complete.

\subsection{Gamma-Ray Bursts (GRBs)}\label{sec:grbs}

One possibility is that \hbox{CDF-S XT1} is connected with a GRB
afterglow or a brighter GRB flare on top of an otherwise standard GRB
afterglow. GRB emission is characterized by time-scales of $\sim$20\,s
for long-duration bursts and $\sim$0.2\,s for short-duration bursts
\citep[hereafter lGRBs and sGRBs, respectively;][]{Meegan1996a}.
Although many questions remain, the commonly accepted lGRB model is
that of a relativistically expanding fireball with associated internal
and external shocks \citep{Meszaros1997a}. After generating the
$\gamma$-ray emission, the expanding fireball shocks the surrounding
material, producing a broadband \hbox{X-ray}-to-radio afterglow that
decays in time as $t^{-a}$ with $a\sim1.2\pm0.3$ unless the Doppler
boosting angle of the decelerating fireball exceeds the opening angle
of the associated jet, at which point the light curve is expected to
steepen \citep[a so-called ``jet break'';][]{Rhoads1999a,
  Zhang2004a}. Alternatively, the currently favoured sGRB
  progenitor scenario features a compact NS-NS or a NS-BH binary
  merger \citep[e.g.,][]{Eichler1989a, Narayan1992a}, induced by
  angular momentum and energy losses due to GW radiation resulting in
  a GW burst \cite[e.g.,][]{Abbott2016a}.  The NS-NS case could
  produce either a millisecond magnetar \cite[e.g.,][]{Zhang2013a} or
  a BH surrounded by a hyper-accreting debris disk, while the NS-BH
  case should yield a larger BH with or without a debris disk,
  depending on whether the NS was tidally disrupted outside of the BH
  event horizon. When a debris disk is present, the combination of the
  high accretion rate and rapid rotation can lead to energy extraction
  via either neutrino-antineutrino annihilation or magnetohydrodynamic
  processes \citep[e.g.,][]{Blandford1977a, Rosswog2002a, Lee2007a},
  which in turn can drive a collimated relativistic outflow. The
  accretion event should also produce more isotropic thermal,
  supernova-like emission on timescales of $\sim$10$^{4}$--10$^{6}$\,s
  known as a 'kilonova'
  \citep[e.g.][]{Metzger2016a,Sun2017a}. Unfortunately, with only a
  few dozen well-characterized SGRBs to date, the parameter range of
  possible properties remains rather open.

To extend the high-energy data available on the transient, we searched
for a possible $\gamma$-ray counterpart in the {\it{Swift}} and
{\it{Fermi}} archives. Unfortunately, neither satellite had coverage in
the direction of the {\it{Chandra}} transient in the few hours
surrounding \hbox{CDF-S XT1} (D. Palmer, H. Krimm, E. {G{\"o}{\u g}{\"u}{\c s},
Y. Kaneko, A. J. van der Horst, private communications), and thus it
is not well-constrained above 10 keV. The field was covered by the
Interplanetary Network \citep{Atteia1987a}, although no counterpart
was detected with a fluence above $10^{-6}~{\rm erg~cm}^{-2}$ and a
peak flux limit of above 1~photon~cm$^{-2}$~s$^{-1}$, both in the
25--150 keV range (K. Hurley, private communication), which excludes
any association with a strong GRB but fails to exclude a faint GRB or
orphan afterglow \citep{Yamazaki2002a, Ghirlanda2015a}.

For comparison, we retrieved the \hbox{X-ray} light curves of
$\sim$760 {\it{Swift}} GRBs with detected \hbox{X-ray} afterglows from
the {\it{Swift}} Burst Analyser \citep{Evans2010a}. Identical to
\citet{Schulze2014a}, we resampled these light curves on a grid
defined by the observed range of \hbox{X-ray} brightnesses and the
timespan probed by the data. If no data were available at a particular
time, we interpolated between adjacent data points (but do not
extrapolate). Figure~\ref{fig:xraygrb} presents the light curve of
\hbox{CDF-S XT1} compared to this {\it{Swift}} GRB distribution in
greyscale.

\begin{figure*}
\vspace{-0.0in}
\begin{center}
\hglue-0.2cm{\includegraphics[height=3.0in, angle=0]{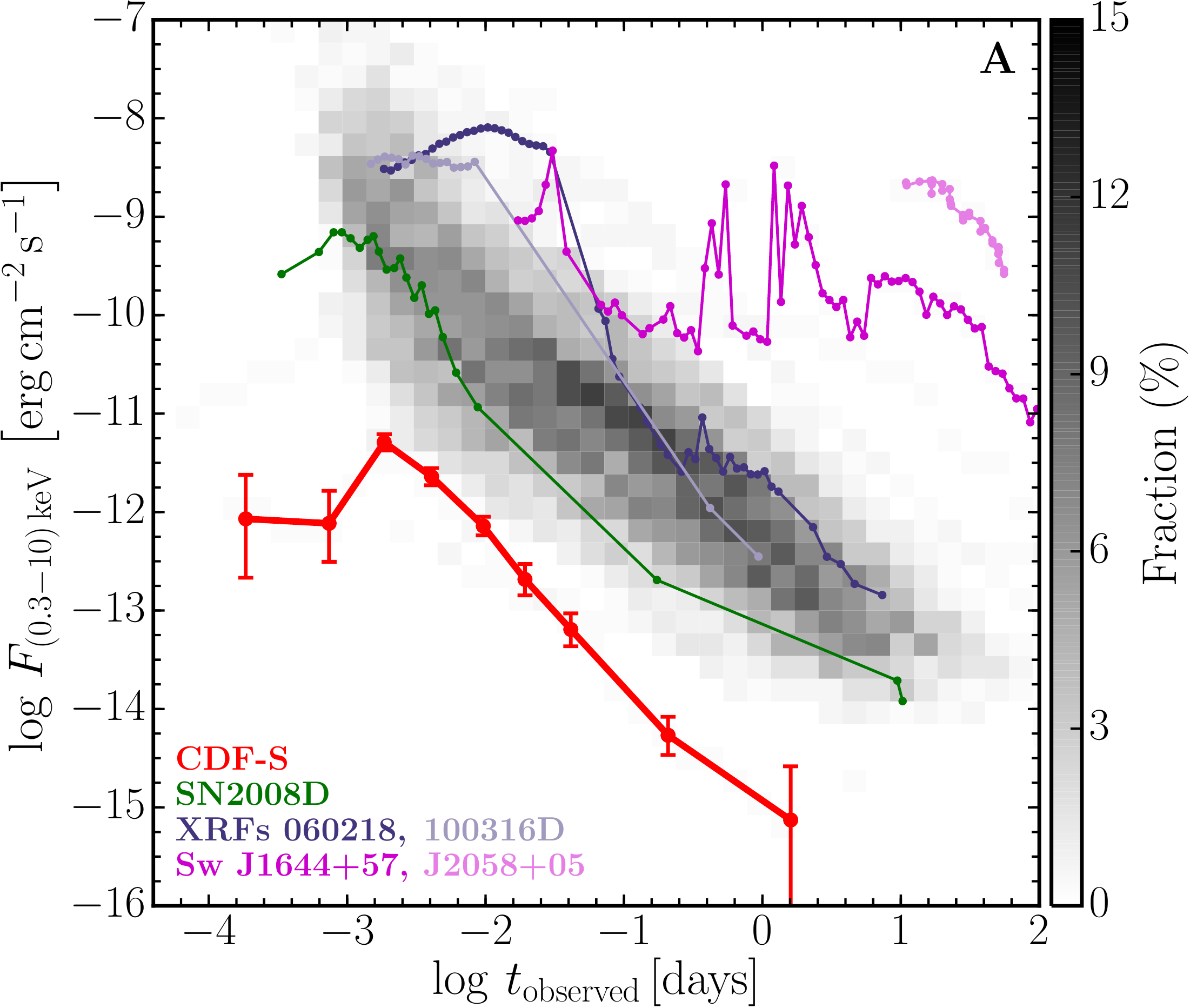}}\hfill
\hglue-1.0cm{\includegraphics[height=3.0in, angle=0]{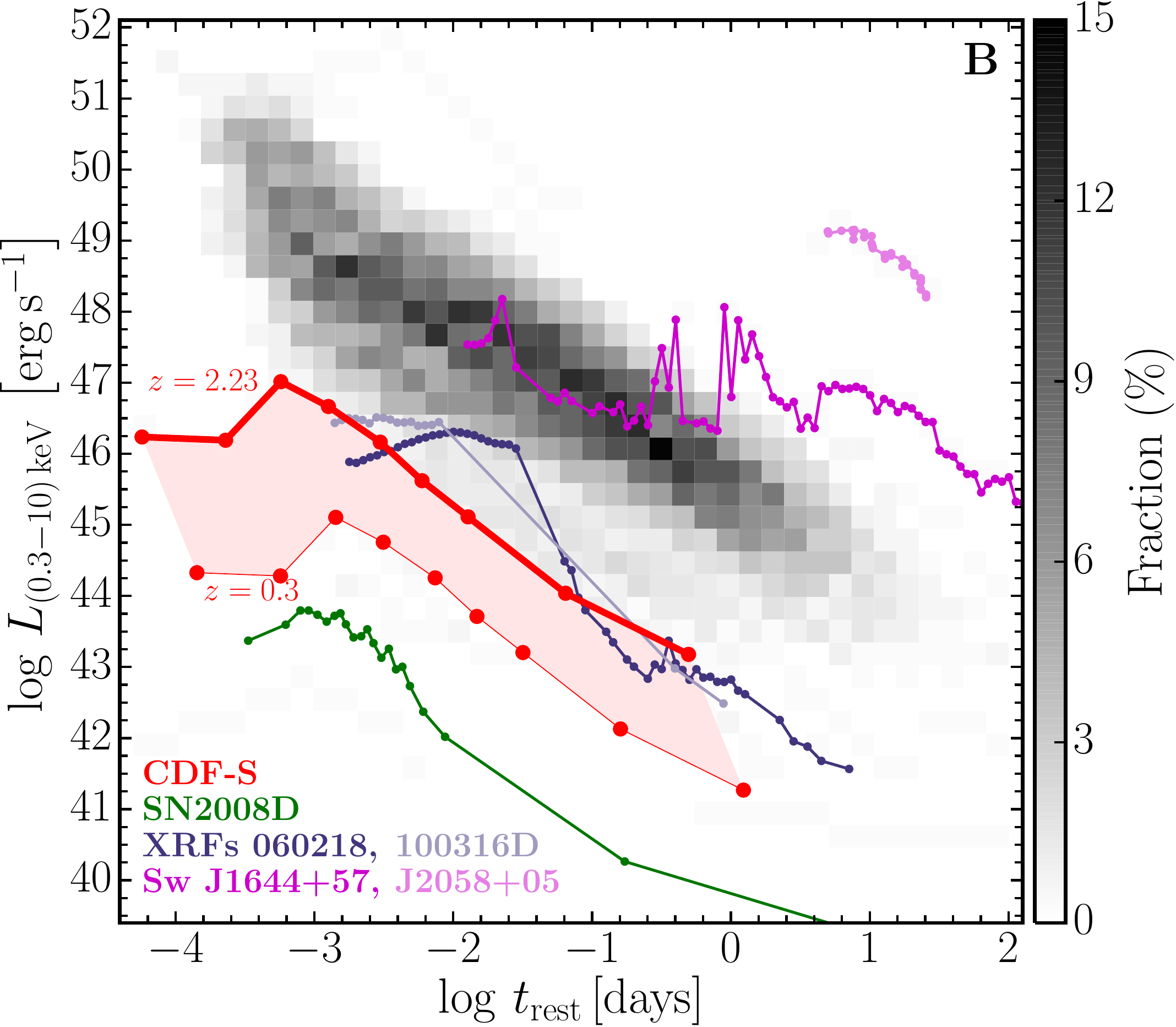}}
\end{center}
\vspace{0in}
\caption{0.3--10\,keV light curve of \hbox{CDF-S XT1}, shown in (A)
  flux and (B) rest-frame luminosity as red points and curves
  (assuming a range of possible redshifts on the luminosity side).
  Shown in grey are the fractional 2D histogram distributions of
  \hbox{X-ray} light curves for $\approx$760 {\it Swift}/BAT detected
  GRBs with detected \hbox{X-ray} afterglows. The power-law decay time
  slope of \hbox{CDF-S XT1} ($a$$=$$-1.53\pm0.27$) is marginally
  steeper than that of the typical GRB afterglow
  \citep[$a$$\approx$$-1.2$;][]{Evans2009a, Racusin2009a}, while its
  luminosity lies at the lower bound of the GRB afterglow
  distribution. The $\approx$100$(1+z)^{-1}$\,s rise time of
  \hbox{CDF-S XT1}, however, is uncharacteristic of GRBs.  A few
  exceptional individual XRF/SNe and beamed TDEs are also
  shown. Low-luminosity XRFs\,080109/SN\,2008D (27 Mpc),
  060218/SN\,2006aj (145 Mpc), and 100316D/SN\,2010bh (263 Mpc) all
  are proposed to have SBO-driven origins \citep{Campana2006a,
    Soderberg2008a, Modjaz2009a, Starling2011a, BarniolDuran2015a} and
  lie in a similar luminosity range as \hbox{CDF-S XT1}, although the
  latter two show distinct plataeus in their early light curves that
  are not observed in \hbox{CDF-S XT1}.  Relativistically beamed TDEs
  \hbox{{\it Swift}\ J1644$+$57} ($z$$=$0.353) and \hbox{{\it
      Swift}\ J2058$+$0516} ($z$$=$1.185) show similar decay time
  slopes over portions of their light curves, but peak much later and
  exhibit significant variability \citep{Bloom2011a, Cenko2012a}.  }
\label{fig:xraygrb}
\end{figure*}

\begin{figure*}
\vspace{-0.0in}
\begin{center}
\hglue-0.2cm{\includegraphics[height=3.0in, angle=0]{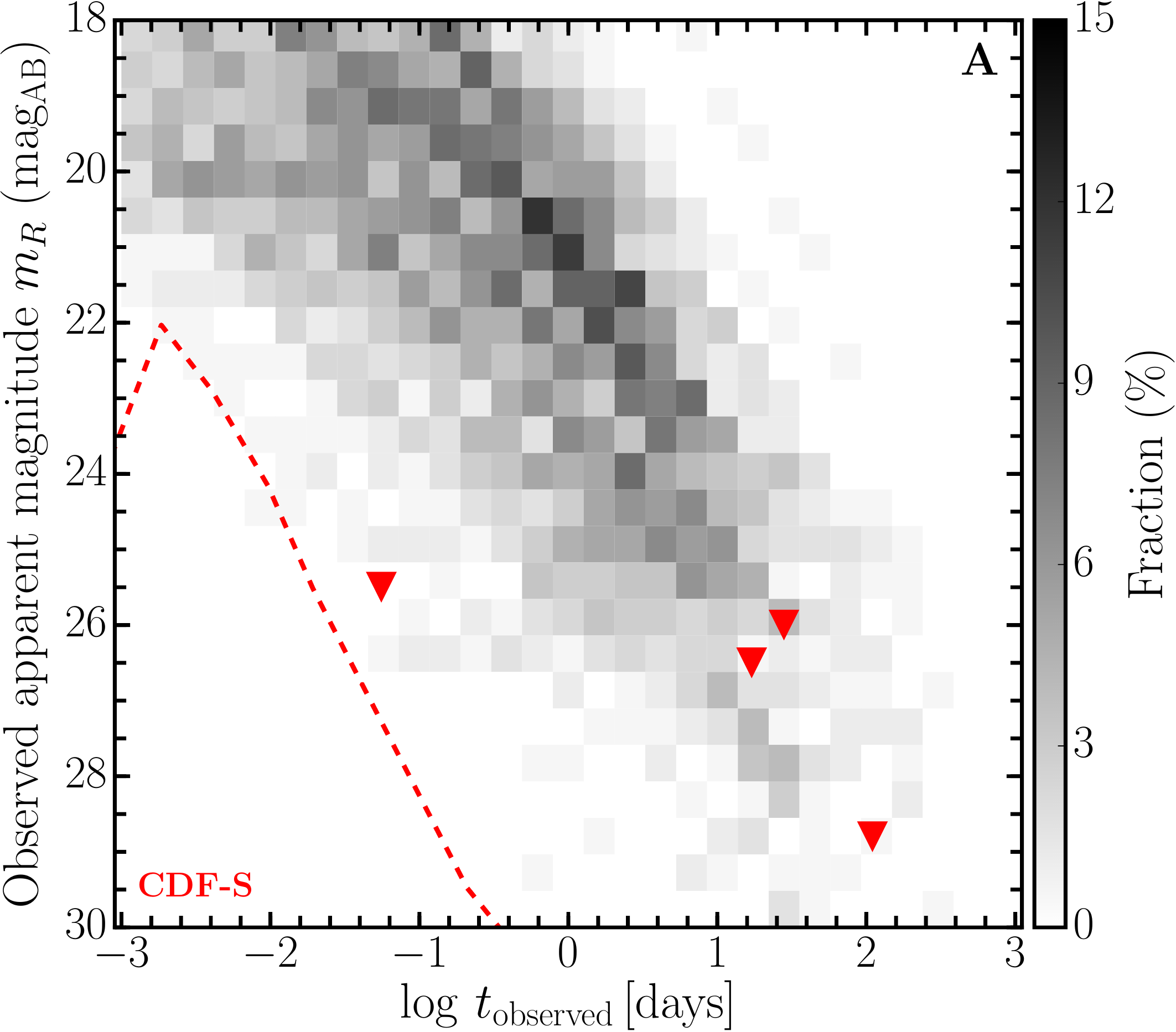}}\hfill
\hglue-1.0cm{\includegraphics[height=3.0in, angle=0]{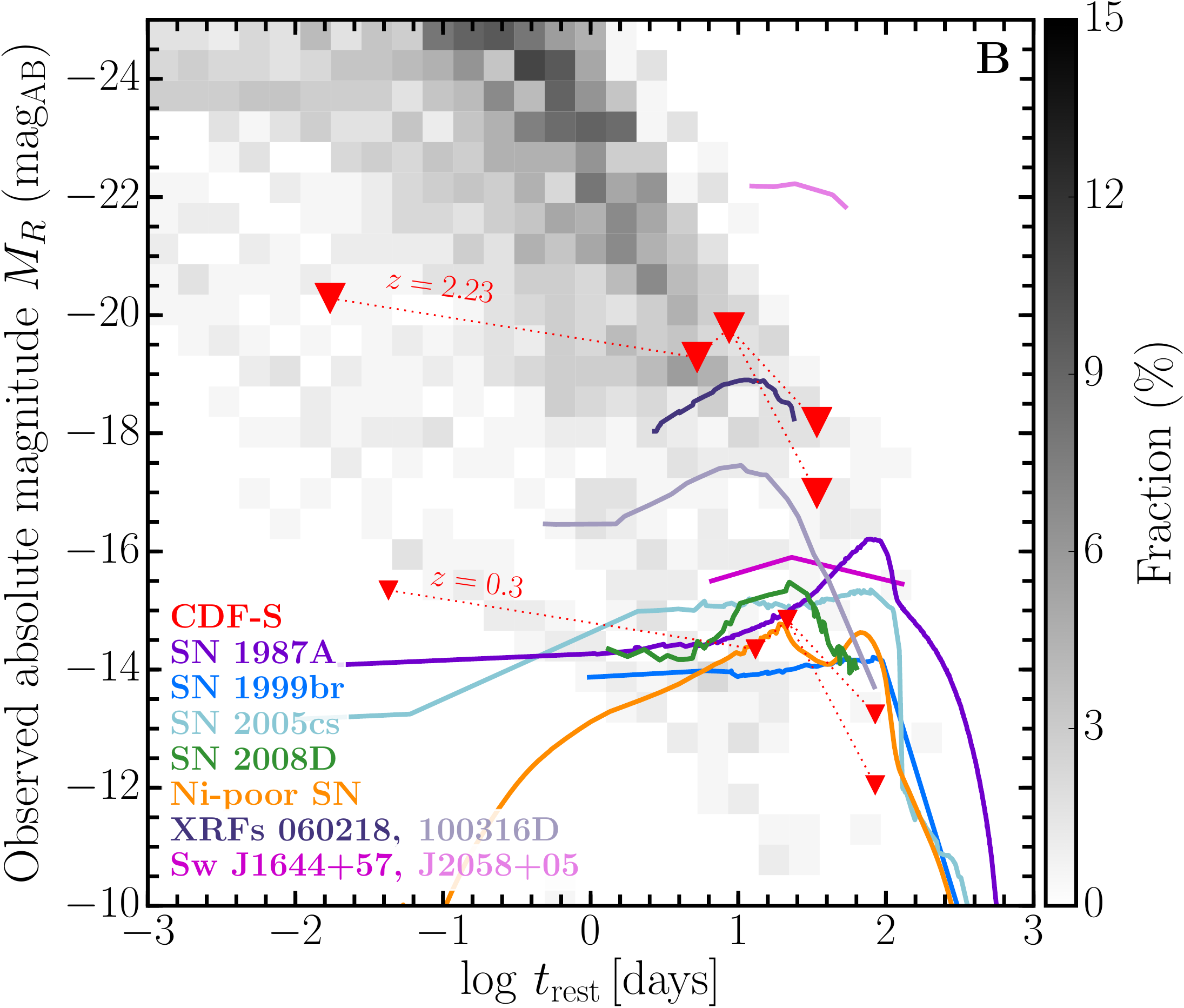}}
\end{center}
\vspace{0in}
\caption{$R$-band (2$\sigma$) upper limits for \hbox{CDF-S XT1}, shown
  in (A) apparent and (B) absolute magnitude as solid red triangles
  connected with red dotted lines (assuming a range of possible
  redshifts on the absolute magnitude side).  The {\it{HST}} $F110W$
  constraint at 111 days is displayed for the range $m_{\rm R}$$-$$m_{\rm
    F110W}$$\approx$0.4--1.0\,mag typical of GRBs and SNe.  Shown in
  grey are the fractional 2D histogram distributions of $R$-band
  afterglow light curves for 166 GRBs with known redshifts, all
  corrected for Galactic extinction. The limits lie at the low end of
  the GRB afterglow optical distribution, with the first limit
  providing the strongest constraint. The dashed red curve in (A)
  shows an extrapolation of the observed \hbox{X-ray} light curve in
  Fig.~\ref{fig:xraygrb} assuming a $\nu^{-0.43}$ power-law spectrum
  between the \hbox{X-ray} and optical regimes; the earliest limit is
  consistent with this, but effectively excludes any strong emission
  above such an extrapolation. A few exceptional individual GRBs, SNe,
  and beamed TDEs are also shown in (B). Low-luminosity events like
  XRF\,080109/SN\,2008D (27 Mpc), XRF\,060218/SN\,2006aj (145 Mpc) and
  XRF\,100316D/SN\,2010bh (263 Mpc) can be excluded out to redshifts
  of $\approx$0.5, $\approx$2, and $\approx$1, respectively, while
  relativistically beamed TDEs like {\it Swift}\,J1644$+$57
  ($z$$=$0.353) and {\it Swift}\,J2058$+$0516 ($z$$=$1.185) are excluded
  out to high redshift. Below $z\approx0.5$--1.0, the {\it{HST}} limit
  excludes even traditionally sub-luminous or late-peaking SNe such as
  Type II-pec SN\,1987A (50 kpc), which rises by $\sim$2 mag to peak
  quite late, Type IIP-pec SN\,1999br (7.1 Mpc) and Type II-P
  SN\,2005cs (8.6 Mpc). These three SNe are among the faintest CCSNe
  known \citep{Richardson2014a}. In some cases, their faintness may be
  related to being Ni-poor CCSNe with strong fallback, for which a
  theoretical light curve is also shown (adopting $M_{\rm
    ZAMS}=25~M_\odot$, $M_{\rm Ni}=0.02$~$M_{\odot}$).  }
\label{fig:optgrb}
\end{figure*}

The peak \hbox{X-ray} flux and full \hbox{X-ray} light curve of
\hbox{CDF-S XT1} are fainter than almost any known GRB \hbox{X-ray}
afterglow \citep{Dereli2015a}. Thus \hbox{CDF-S XT1} would need to be
an intrinsically low-luminosity, misaligned, or high-redshift GRB. The
light curve decay time slope of \hbox{CDF-S XT1}
($a$$=$$-1.53\pm0.27$) appears fairly constant and marginally steeper
than the median afterglow decay time slope for GRBs
\citep[$a$$\approx$-1.2;][]{Evans2009a, Racusin2009a}, while its
\hbox{X-ray} spectral slope ($\Gamma$$=$1.43$^{+0.26}_{-0.15}$) lies
in the hardest $\approx$10\% of the standard afterglow distribution
over comparable energy bands
\citep[$\Gamma$$\approx$1.70$\pm$0.15;][]{Wang2015a}. While few GRBs
have been well-characterized below 10\,keV as they initially exploded,
a substantial subset of lGRBs have been observed within 10--100\,s of
the prompt burst, at which point the low-energy tail of the prompt
emission has been seen \citep{Tagliaferri2005a, Barthelmy2005a}; this
feature is likely responsible for the steeper initial decline seen in
the greyscale histogram distributions of {\it Swift}-detected GRBs
shown in Fig.~\ref{fig:xraygrb}.\footnote{We note that a handful of
  {\it Swift}-detected GRBs do show initial rises and peaks around
  100--500\,s, similar to \hbox{CDF-S XT1}, but their subsequent
  behaviour appears quite distinct from that of \hbox{CDF-S XT1}, with
  multiple strong flares and clear breaks.}  Early observational
constraints of sGRBs are far more difficult to obtain due to their
limited durations, although they also are expected to show early
contributions from the prompt emission \citep{Villasenor2005a}. Thus a
critical discriminator in \hbox{CDF-S XT1}'s \hbox{X-ray} light curve
is its $\approx$100$(1$$+$$z)^{-1}$\,s rest-frame rise time (where $z$
is redshift), which contrasts sharply with the strong early emission
and spectral softening expected from both lGRBs and
sGRBs. Any burst of photons associated with the prompt
emission would have been detected easily by {\it Chandra}, if they
extended below 10--20\,keV in the rest-frame.

During the first several hours after a GRB, the \hbox{X-ray} afterglow
is frequently characterized by flaring episodes
\citep{Chincarini2007a, Chincarini2010a, Margutti2011a}. The peak time
of \hbox{CDF-S XT1} is consistent with that for GRB flares, although
the spectral slope and decay time slope are on the hard and slow ends
of their respective distributions. The duration of \hbox{CDF-S XT1},
however, is substantially longer than those seen in GRB flares (i.e.,
$\sim$10-300\,s), and thus is unlikely to fit cleanly into such a
scenario.

In the optical, we compare to composite $R$-band light curves derived
from a database of optical and NIR measurements of 166 GRBs with known
redshifts \citep{Kann2006a, Kann2010a, Kann2011a, Nicuesa2012a}. These
data were gridded in an identical manner to the \hbox{X-ray} data;
Figure~\ref{fig:optgrb} shows the corresponding density distribution
(cropped at a lower limit of 10$^{-3}$ days, since that is the regime
most relevant for our observed contraints and allows better
visualization of the ``busy'' 10--100 day region). Comparing the
initial $R$-band limit ($m_{\rm R}$$\gtrsim$25.7\,mag at
$\approx$80\,minutes post-transient) to this optical density
distribution, the transient is again fainter than $\gtrsim$99\% of
known GRB afterglows. As seen in Fig.~\ref{fig:optgrb}, the early
2$\sigma$ limit lies only $\sim$1 magnitude above a $\nu^{-0.43}$
power-law extrapolation of the \hbox{X-ray} light curve into the
optical band (dashed red curve in Fig.~\ref{fig:optgrb}), thereby
providing a relatively stringent constraint on any excess emission
above this estimate. Given the prompt \hbox{X-ray} emission, the early
limit rules out a standard off-axis jet scenario, wherein we would
expect to find a relatively normal, bright optical GRB afterglow
associated with weak \hbox{X-ray} emission \citep{vanEerten2010a,
  vanEerten2011a}. In fact, based on synchrotron closure relations
between the \hbox{X-ray} and optical emission, a large fraction of
parameter space can be excluded. Thus to explain the observations in a GRB
scenario, the transient would need to be intrinsically low-luminosity,
reddened by at least a few magnitudes, and/or at redshift
$z$$\gtrsim$3.6--5.0 (and hence not associated with the apparent
host).

Each of these possibilities shares a relatively low probability
\citep{Jakobsson2012a, Covino2013a}. The strong association with the
$z_{\rm ph}$$\sim$2.2 host galaxy appears to rule out the high
redshift option and lowers the probability of the low-luminosity
option. Alternatively, a small fraction of sGRBs show extremely weak
optical emission, as seen in Fig.~\ref{fig:optgrb}, allowing an
off-axis sGRB with weak optical emission to remain a viable option
\citep{Lazzati2016a}.

To put the radio limits into context with radio GRB afterglows, we
compared the reported limits listed previously to the work of
\citet{Chandra2012a}. In particular, the ATCA 9\,GHz limit implies a
faint afterglow, although a significant number of GRBs have evaded
detection with deeper observations.

Based on the above considerations, the \hbox{X-ray} transient does not
appear to be fully consistent with the properties of most known GRBs,
nor the predictions for off-axis or weaker ones. The relatively low
X-ray fluxes exclude all but low-luminosity, off-axis, or
$z$$\gtrsim$4 GRB solutions, while the optical transient limits
further exclude most standard off-axis GRB solutions and necessitate
weak or absorbed optical emission. The strong association with a
$z_{\rm ph}$$\sim$2.2 host galaxy appears to exclude the high-z
solution. The lack of prompt emission below $\sim$20\,keV rest-frame
further excludes any on-axis / strongly beamed scenario. Taken
together, only a few tentative options may remain. 

One is an ``orphan'' off-axis sGRB, which furthermore must have
exceptionally weak optical and radio emission. Based on predictions,
such objects could exist, although none has yet been confirmed.
Models of compact object mergers \citep[e.g.,][]{Metzger2014a,
  Sun2017a} suggest that the initial X-ray light curves are likely to
be optically thick and ``turn-on'' over timescales of many hours to
days as it expands, with peak X-ray luminosities in the range of
$\sim$$10^{44}$--$10^{46}$\,erg\,s$^{-1}$ followed by a $\sim$t$^{-2}$
decay. While such peak luminosities are naively compatible with
\hbox{CDF-S XT1}, the peak times are 2--3 dex longer. Among the many
X-ray models of \citet{Sun2017a}, some allow for for earlier turn-ons,
but with correspondingly higher peak X-ray luminosities. Some compact
object merger models \citep[e.g.,][]{Metzger2014a} are additionally
expected produce strong optical/near-IR emission, which we do not
see. Considerable fine-tuning and/or revision of sGRB models may be
required in order to more satisfactorily match the observational
constraints of \hbox{CDF-S XT1}.

Another possibility is an explanation as a low-luminosity GRB at
$z$$\gtrsim$2, although the X-ray light curve and spectral properties
of \hbox{CDF-S XT1} remain distinct from the best-studied
low-luminosity GRBs.

\subsection{Shock Breakout (SBO)}\label{sec:sbo}

One intriguing possibility is that the \hbox{X-ray} transient
represents the SBO from a core-collapse supernova (CCSN). An initial
flash of thermal UV (or soft \hbox{X-ray}) radiation is expected when
the CCSN shock wave emerges from the stellar surface of the progenitor
\citep{Falk1977a, Klein1978a, Matzner1999a, Schawinski2008a,
  Ganot2016a}. The character of the SBO depends primarily on the
density structure of the progenitor and the explosion energy driving
the shock \citep{Chevalier2011a, Gezari2015a}, resulting in SBOs with
expected initial temperatures of $\sim$$10^{5}$ to
5$\times$10$^{6}$\,K and durations of $\sim$100--5000\,s. The typical
bolometric luminosity associated with an SBO is generally of order
$\sim$10$^{44}$--10$^{45}$\,erg\,s$^{-1}$ \citep{Ensman1992a,
  Tominaga2011a}, while the emission in a given \hbox{X-ray} band
(e.g., 0.3--10 or 2--10\,keV) will be less ($\sim$1--87\% for quoted
temperature range). Moreover, a sufficiently compact progenitor with
an energetic explosion could produce relativistic effects that
substantially harden the \hbox{X-ray} spectrum and lead to significant
$>1$\,keV emission, although only for a relatively short time
(1--100\,s) and with dramatic spectral and temporal evolution
\citep{Tolstov2013a}. After the SBO, the outer layers of the star
should enter an adiabatic expansion and cooling phase for $\sim$1--2
days, followed by a plateau phase thereafter as radiative diffusion
takes over \citep{Chevalier1992a, Popov1993a}. If the stellar wind or
circumstellar material (CSM) is sufficiently dense, it could intensify
or prolong the SBO (by up to factors of $\sim$10) and delay the
subsequent phases \citep{Moriya2011a, Balberg2011a, Svirski2012a}.

While the duration of the \hbox{X-ray} transient is consistent with
that of longer SBOs, the hard observed spectrum (e.g., $kT$$>$5\,keV
at 3$\sigma$) appears inconsistent with the relatively low expected
temperatures ($kT$$\sim$0.01--1\,keV) for such typical SBOs. For a
fixed total energy budget from a SNe, there should be a tradeoff
between luminosity and temperature, whereby a larger radius at which
the SBO occurs could give a higher luminosity, but a lower blackbody
temperature. This makes an SBO interpretation hard to satisfy with the
observed properties. The most promising models are explosions of blue
supergiants like SN\,1987A, which can achieve bolometric luminosities
as high as $10^{45}$\,erg s$^{-1}$ and SEDs peaking at $\sim$10\,keV,
however only for durations of $\sim$100\,s (A. Tolstov, private
communication).  Alternatively, a relativistic scenario might be able
to explain the observed \hbox{X-ray} photon index, but we do not
observe the characteristic strong spectral and temporal evolution (a
power-law decay time slope of $a>2$). Additionally, by $z\sim0.4$ the
peak \hbox{X-ray} luminosity of the transient already exceeds the
predicted bolometric peak luminosity for a SBO associated with a
$E_{\rm explosion}$$\approx$$10^{51}$\,erg progenitor. Thus the SBO
scenario is only viable at low redshift, which remains possible but
unlikely based on the photometric redshift probability
distribution. To accommodate the best-fitting redshift of $z$$=$2.23
with a SBO scenario requires an explosion energy of $\gtrsim$10$^{52}$
erg and/or an optically thick CSM. 

The best-studied SBO candidate to date is \hbox{X-ray} Flash (XRF)
080109/SN\,2008D (27 Mpc), which compared to \hbox{CDF-S XT1} has a
mildly different \hbox{X-ray} light curve evolution (more gradual
rise, broader ``peak'', and broken decline) but much softer
\hbox{X-ray} spectrum \citep[with $kT$$\sim$0.7\,keV or
  $\Gamma$$\sim$2.1;][]{Soderberg2008a, Modjaz2009a}. We can also
compare to the XRFs 031203/SN\,2003lw (475\,Mpc), 060218/SN\,2006aj
(145 Mpc) and 100316D/SN\,2010bh (263 Mpc), which are also proposed to
have SBO-driven origins. While the high-energy ($>$2\,keV)
\hbox{X-ray} spectral slopes over comparable bands are consistent, the
latter two XRFs show significant soft thermal components
($kT$$\sim$0.1--0.2\,keV) which become dominant beyond $\sim$1000\,s
\citep{Campana2006a, Starling2011a, BarniolDuran2015a}, while
\hbox{CDF-S XT1} appears to marginally harden at late
times.\footnote{XRF 031203/SN\,2003lw was observed in the pre-{\it
    Swift} era and hence has substantially sparser and later X-ray and
  optical follow-up constraints. Notably, the portions of the X-ray
  and optical light curves that are well sampled appear similar to
  those of XRF 060218/SN\,2006aj \citep{Watson2004a, Mazzali2006a}. As
  such, we do not include it in Figs.~\ref{fig:xraygrb} and
  \ref{fig:optgrb} for clarity.} In addition, the early \hbox{X-ray}
light curves of these two events lie in stark contrast to \hbox{CDF-S
  XT1}, as they are substantially flatter and longer lasting, with
steeper late time declines \citep{Campana2006a, Starling2011a,
  BarniolDuran2015a}.

Another critical aspect of the SBO scenario is, of course, the
expectation of subsequent strong UV/optical emission associated with
the standard CCSN light curve. Our combined optical/NIR constraints,
shown in Fig.~\ref{fig:optgrb}, appear to rule out several of the
faintest known SNe light curves \citep{Richardson2014a} if placed
closer than $z$$\sim$0.5, and more luminous ones out to
$z$$\lesssim$1--2, with the most critical constraints arising from the
initial VIMOS and latest {\it{HST}} data points. Adopting the
\hbox{X-ray} to optical flux ratios of XRFs 080109/SN\,2008D,
031203/SN\,2003lw, 060218/SN\,2006aj and 100316D/SN\,2010bh, the
associated SNe light curves would also all have been easily
detected. The most sub-luminous SNe are thought to have extremely low
nickel yields \citep[i.e., nickel masses of
  $\approx$0.002--0.075\,$M_{\odot}$;][]{Hamuy2003a}, suggesting that
if this event were an SBO, it would potentially require little nickel
production, and consequently strong fallback. We also show the
theoretical light curve for a Ni-poor core-collapse SN with strong
fallback, from a ZAMS $M$$=$25\,$M_\odot$ progenitor and a nickel mass
of 0.02\,$M_{\odot}$. To evade the optical/NIR limits would require a
rather contrived scenario of nearly complete fallback. Alternatively,
if we redden the comparison light curves by $A_{\rm
  R}$$=$0.3--1.3\,mag, they can fit the early ground-based limits,
although all are still strongly excluded by the NIR {\it{HST}} limit
at high significance unless significantly stronger reddening is
assumed.

\subsection{Tidal Disruption Event (TDE)}\label{sec:tde}

A further possibility is that the transient was a TDE.  TDEs occur
when a star passes exceptionally close to a $\gtrsim$$10^{4}$
$M_{\odot}$ BH \citep{Rees1988a, Phinney1989a,
  Burrows2011a}, such that it experiences tidal forces which exceed
its self-gravity, allowing the star to be shredded. Luminous thermal
emission at soft \hbox{X-ray} through optical wavelengths is generated either
by the accretion of this gas onto the BH [often limited to $L_{\rm
    Bol}$$\approx$$L_{\rm Edd}$$\approx$$1.3\times10^{44}$
  erg\,s$^{-1}$ ($M_{\rm BH}/10^{6}$ $M_{\odot}$)] and/or the initial
shocks due to colliding stellar debris streams
\citep{Guillochon2015a}.  The tidal disruption radius is given by
$r_{\rm TDE}$ $\approx$ $(2M_{\rm BH}/m_{*})^{1/3} R_{*}$, where
$M_{\rm BH}$ is the mass of the BH, while $m_{*}$ and $R_{*}$ are the
mass and radius of the star, respectively. This radius effectively
dictates in what band the thermal radiation will peak, with the
effective temperature given by $T_{\rm
  eff}$$\approx$2.5$\times$$10^{5}$\,K\, ($M_{\rm BH}/10^{6}$
$M_{\odot}$)$^{1/12}$ ($R_{*}/R_{\odot}$)$^{-1/2}$
($m_{*}/M_{\odot}$)$^{-1/6}$. For normal main-sequence stars disrupted
around $10^{6}$--$10^{8}$ $M_{\odot}$ BHs, the radiation should peak
between $T_{\rm eff}$$\sim$10$^{4}$--10$^{6}$\,K. The time-scale for
the emission to rise to maximum is given by $t_{\rm
  min}\approx$0.11\,yr\,($M_{\rm BH}/10^{6}$$M_{\odot}$)$^{1/2}$
($R_{*}/R_{\odot}$)$^{-3/2}$ ($m_{*}/M_{\odot}$)$^{-1}$, after which
the bolometric light curve is predicted to follow a $\approx t^{-5/3}$
power-law decay. In rare cases, material accreting onto the BH may
produce a relativistic jet which gives rise to non-thermal
$\gamma$-ray, \hbox{X-ray} and radio emission which can appear orders of
magnitude more luminous and can be strongly variable
\citep[e.g.,][]{Burrows2011a, Bloom2011a, Cenko2012a}.

The decay time slope of the transient is fully consistent with the
predictions for TDEs. However, the fast rise time and hard
\hbox{X-ray} flux of \hbox{CDF-S XT1} strongly exclude all ``normal''
stars and supermassive BHs ($>$$10^{6}$\,$M_{\odot}$). The only viable
remaining parameter space is for a TDE comprised of a white dwarf (WD:
0.008--0.02$R_{\odot}$, 1\,$M_{\odot}$) and an intermediate-mass BH
(IMBH; $\sim$10$^{3}$--10$^{4}$\,$M_{\odot}$), although even in such a
scenario it may be difficult to explain the hard observed \hbox{X-ray}
spectral slope.  Furthermore, the resulting Eddington luminosity for
this TDE scenario would be at least two orders of magnitude too low
compared to what is expected for the redshift range of the associated
host galaxy (Table~\ref{tab:energetics}). One alternative could be
that the emission arises from a relativistic jet produced by the TDE,
although then we might expect substantially stronger variability
fluctuations than is observed from the relatively smooth power-law
decay of the transient's \hbox{X-ray} light curve
\citep{Levan2011a}. Moreover, the ratio of \hbox{X-ray} emission to
the optical and radio limits is $\gtrsim100$ times larger than those
from the beamed TDEs {\it Swift}\,J1644$+$57 and {\it
  Swift}\,J2058$+$0516 \citep{Bloom2011a, Cenko2012a}. Moreover, the
beaming requirements become quite extreme with increasing redshift.
Thus, relativistically beamed emission from a TDE comprised of a WD
and an IMBH remains only a remote possibility.

\subsection{Galactic Origin}\label{sec:galactic}

There are at least some similarities between the reported
characteristics of \hbox{CDF-S XT1} and a wide variety of \hbox{X-ray}
emitting Galactic phenomena. We limit the discussion here only to the
possibilities which have similar \hbox{X-ray} transient time-scales and
are unlikely to require bright optical or NIR counterparts, as these
are easily excluded by our imaging and line-of-sight through the
Galaxy (for instance, this removes most high and low-mass \hbox{X-ray}
binary systems).

One remaining possibility is an origin as an M-dwarf or brown-dwarf
flare.  Magnetically-active dwarfs ($\sim$30\% of M dwarfs, $\sim$5\%
of brown dwarfs) are known to flare on time-scales from minutes to
hours, exhibiting flux increases by factors of a few to hundreds in
the radio, optical blue, UV, and/or soft \hbox{X-ray}
\citep{Schmitt2004a, Mitra2005a, Berger2006a, Welsh2007a}.  The flares
can be short ``compact'' ($L$$\lesssim$$10^{30}$ erg\,s$^{-1}$,
$<1$\,h) or ``long'' ($L$$\lesssim$$10^{32}$ erg\,s$^{-1}$,
$\ge1$\,h). Flares are often recurrent on time-scales of hours to
years, and in the \hbox{X-ray} band at least typically have thermal
\hbox{X-ray} spectral signatures with $kT$$=$0.5--1\,keV, both of
which are inconsistent with \hbox{CDF-S XT1}. M dwarfs tend to be more
\hbox{X-ray} active than brown dwarfs \citep{Berger2006a,
  Williams2014a}, with $\log(L_{\rm R}/L_{\rm X})$$\sim$$-$15.5 in
units of $\log{{\rm (Hz}^{-1}{\rm )}}$ for a wide range of stars down
to spectral types of about M7, after which this ratio rapidly climbs
to $\sim$$-$12 for brown dwarfs. Thus, relative to the observed
\hbox{X-ray} peak flux, the radio survey limits mentioned previously
should have been more than sufficient to detect radio flares from a
brown dwarf and most M-dwarfs, if any occurred during the radio
observations. While the best counterpart for the transient, source
\#1, is clearly extended in multiple images, there remains a low
probability ($<$0.3\%) that that the transient could be matched to
source \#2, which is potentially consistent with a $m_{\rm
  R}$$=$27.38\,mag M-dwarf star. Alternatively, an even fainter dwarf
could lie below the {\it HST} detection threshold, although the low
random probability of spatial coincidence with a background galaxy
strongly argues against this. Notably, M dwarfs typically have
absolute magnitudes of $M_{\rm R}$$\sim$8--14\,mag
\citep{Bochanski2011a}, such that source \#2 would have to lie at
$\gtrsim$5--75\,kpc (i.e., in the halo), while an undetected M dwarf
would lie even further away. Similarly, brown dwarfs have typical
absolute magnitudes of $M_{\rm J}$$\sim$15--25 \citep{Tinney2014a},
such that an undetected source must lie at $\gtrsim$30\,pc--3\,kpc.
However, such distances would imply an M-dwarf \hbox{X-ray} luminosity
of $L_{\rm X}$$\gtrsim$(3.4--850)$\times$10$^{35}$ erg\,s$^{-1}$ or a
brown dwarf \hbox{X-ray} luminosity of $L_{\rm
  X}$$\gtrsim$(5.5--54903)$\times$10$^{29}$ erg\,s$^{-1}$, which are
at least factors of $\gtrsim$10$^{3}$--10$^{5}$ larger than typical
flares seen from M dwarfs \citep{Pandey2008a,Pye2015a} and brown
dwarfs \citep{Berger2006a}, respectively. Thus the observed transient
properties appear inconsistent with those of dwarf flares.

Another possibility is that the transient was the result of a magnetar
outburst. Magnetars are spinning-down, isolated neutron stars which
have relatively slow rotation rates ($\sim$1--10\,s) and possess
extremely strong magnetic fields that are considered to power
characteristic and recurrent bursts of \hbox{X-ray}s and $\gamma$-ray
radiation \citep[hence their designations as ``soft gamma repeaters'',
  SGRs, or ``anomalous \hbox{X-ray} pulsars'',
  AXPs;][]{Mereghetti2015a}.  They lack obvious companions from which
to accrete, yet have apparent \hbox{X-ray} luminosities during
outbursts which can often be super-Eddington and reach luminosities as
high as $\sim$10$^{47}$\,erg\,s$^{-1}$ \citep{Palmer2005a}; these
cannot be explained by rotation power alone. Their strong magnetic
fields are predicted to decay on time-scales of $\lesssim$10,000
years, after which their activity ceases.  Among the $\approx$26
magnetars known \citep{Olausen2014a}, many are found near OB
associations and/or SN remnants and all lie in the thin disk of the
Galaxy or the Magellanic clouds, implying magnetars are possibly a
rare by-product of massive O stars.  Roughly half of the magnetars are
persistent \hbox{X-ray} sources with fluxes of
$\sim$10$^{-14}$--10$^{-10}$ erg\,s$^{-1}$\,cm$^{-2}$ (or equivalently
quiescent \hbox{X-ray} luminosities of order
$\sim$$10^{35}$\,erg\,s$^{-1}$). The rest were primarily discovered
during bright, short outbursts (0.1--1.0\,s) or giant flares
(0.5--40\,s), $\sim$10--1000 times brighter than their anticipated
quiescent phases, whose properties still remain relatively poorly
known. The rises and decays of these outburst/flare episodes show
different durations and shapes ($\sim$1\,week to months), but the
decays are generally characterized by a spectral softening. The
outburst duty cycle remains poorly known, as multiple distinct
outburst episodes have only been detected from a few magnetars to
date.  The \hbox{X-ray} spectra are generally fit with two-component
blackbody ($kT_{\rm BB}$$\sim$0.5\,keV) and power-law
($\Gamma$$\sim$1--4) models. A subset of magnetars have optical and
radio counterparts. Based primarily on the strong association with
recent star forming regions and SN remnants (the high Galactic
latitude CDF-S field is far from any known Galactic star-forming
region), as well as the more sporadic and longer duration rise and
decay times expected, \hbox{CDF-S XT1} appears highly unlikely to be a
Galactic magnetar.

Finally, the \hbox{X-ray} properties of \hbox{CDF-S XT1} could be
related to compact object such as an asteroid hitting an isolated
foreground NS \citep{Colgate1981a,vanBuren1981a,Campana2011a}. This
possibility, which was originally suggested to explain GRBs, is
difficult to rule out based on the observational data alone due to the
wide parameter range of transients that can be produced. However, the
combined probability that such an event occurs on a NS which just
happens to align with a faint extragalactic source to better than
1$\sigma$ is quite low ($\ll$0.1\%), given an expected NS number
density out to 30 kpc of $\sim$1000 deg$^{-2}$ \citep{Sartore2010a}
and the source density of $m_{\rm F160W}$$<$27.5\,mag galaxies in the
CANDELS field (0.088 arcsec$^{-2}$).

\section{Event Rates}\label{sec:rates}

Regardless of origin, the fact that this event occurred in a
pencil-beam survey field like the CDF-S naively implies a relatively
high occurrence rate. However, although a handful of high-amplitude,
fast X-ray transients have been reported in the literature to date
\citep{Jonker2013a,Glennie2015a,deLuca2016a}, none appears to have the
X-ray and optical transient properties of \hbox{CDF-S XT1} nor an
association with such a faint optical host. To quantify this, we first
perform a search of the {\it{Chandra}} archive to determine the
frequency of events such as \hbox{CDF-S XT1}, and then estimate their
occurrence rate.

\subsection{Comparable Events}\label{sec:csc_search}

To determine the uniqueness of this transient, we conducted an
archival search for similar variable events. Due to the limited
variability information available (e.g., no easy access to individual
source photon tables) in the most recent {\it{XMM-Newton}} 3XMM DR5
and {\it{Swift}} 1SXPS source catalogs \citep{Rosen2016a,Evans2014a},
we found it infeasible to assess properly whether such a source was
detected by either observatory, and thus only conducted a search for
similar events observed by {\it Chandra} using the Chandra Source
Catalog \citep[CSC v1.17;][]{Evans2010b}. Most critically, the CSC is
the only source catalog that provides easy and straightforward access
to photon event lists and light curves, which we considerd absolutely
essential to characterize the nature of the variability of each
source. Even so, the current version of the CSC only contains
relatively bright sources from the first 11 cycles (up to 2010 August
10), which factors into our rate calculations below.

We began by searching the CSC for all securely variable sources with a
peak flux of $F_{\rm 2-10\,keV}>1\times10^{-12}$ erg\,s$^{-1}$
cm$^{-2}$ (or their count-rate equivalent, adopting the best-fitting
spectral slope of the transient) and which varied in flux by at least
a factor of 10. This should find any similar transients down to a
factor of five weaker than \hbox{CDF-S XT1}, if they exist. Such a
transient should be easily detectable in virtually any {\it{Chandra}}
observation. At this flux, it would also be detectable in
{\it{XMM-Newton}} or {\it{Swift}}/XRT, although it might be difficult
to characterize it in {\it{Swift}}/XRT data due to the typically short
(1--2\,ks) observations this instrument executes. The above critieria
are obviously conservative, as {\it{Chandra}}'s sensitivity could
allow a search up to a factor of 10 deeper. However, with so few
observed counts ($\sim$10 photons), it would be impossible to
determine with much certainty whether the transient truly is similar
to \hbox{CDF-S XT1}.

\begin{table*}
{\scriptsize
\caption{ 
Light curve characterization for 184 candidates from CSC search.
{\it Col. 1}: Light curve category ({\it criterion \#2}).
{\it Col. 2}: Number found.
{\it Col. 3}: Variability characteristics.
{\it Col. 4}: Number within $|b|$$\le$10$^{\circ}$ (i.e., considered Galactic).
{\it Col. 5}: Number with clear optical counterparts or associations with globular clusters, star forming regions, or galaxies  ({\it criterion \#1}).
{\it Col. 6}: Number that exhibit a best-fitting \hbox{X-ray} spectral slope of $\Gamma$$<$2.0. In many cases, the spectra and hardness ratios are not of sufficient quality to robustly determine a spectral slope. Therefore we provide the possible range.
({\it criterion \#3}).
\label{tab:CSCsearch}}
\centering
\begin{tabular}{l|r|l|c|c|c}
\hline
Cat. & \# & Var. type & GalPlane? & Opt? & Hard? \\
\hline
1 & 129 & \parbox[l]{12cm}{recurrent/persistent transients (flaring, eclipsing, gradual, etc.).} & 20 & 123 & 14--47 \\
\hline
2 & 29 & \parbox[l]{12cm}{non-recurrent transients with marginal or no prior/post detections; exhibit $\gtrsim 4$\,ks rises and/or decay time-scales several times longer than \hbox{CDF-S XT1}.} & 7  & 29 & 12  \\
\hline
3 & 7 & \parbox[l]{12cm}{non-recurrent transients with no prior/post detections; exhibit 2--4\,ks rise times, decay time-scales several times longer than \hbox{CDF-S XT1} and/or flattening either early or late in the transient decay.}  & 3 & 7 & 4  \\
\hline
4 &  5 & \parbox[l]{12cm}{non-recurrent transients with no prior/post detections similar to Cat 3, but time series terminates or is of insufficient statistical quality to assess further.}  & 2 & 4 & 2--4  \\
\hline
5 &  4 & \parbox[l]{12cm}{non-recurrent transients with no prior/post detections and show consistent \hbox{X-ray} light curve shapes with peak of $\lesssim$10$^{3}$.} & 2 & 4 & 2  \\
\hline
6 &  6 & \parbox[l]{12cm}{non-recurrent transients potentially similar to Cat 5, but have significant low-level activity in the $\sim$10$^{3}$--10$^{4.5}$\ s prior to Cat 5 light curve shapes.} & 4 & 5 & 1--3 \\
\hline
7 &  4 &  \parbox[l]{12cm}{non-recurrent transients with no prior/post detections; exhibit weak short rise and fall behavior within 100--500\,s and dominated by noise at late times.} & 1 & 3 & 0--4 \\
\hline
\end{tabular}}
\end{table*}

To select sources similar to \hbox{CDF-S XT1}, we adopted the
following CSC parameters for sources observed with the ACIS-I and
ACIS-S detectors, as well as the High Resolution Camera (HRC):
variability probability $var\_prob$$>$0.9; maximum variability count
rate $o.var\_max$$>$0.06 cnt/s; and minimum variability count rate
$o.var\_min$$<$0.006 cnts/s.
We found 184 unique matches.
Of these, 39 lie within $|b|$$<$10$^{\circ}$ and are presumably Galactic in
nature, while at a minimum a further 19 and 82 can be spatially
associated with Milky Way globular clusters and young star-forming
regions outside the Galactic plane ($|b|$$>$10$^{\circ}$), respectively.
Based on the {\it{Chandra}} positional errors, 149 candidates have
Digitized Sky Survey (DSS), Two Micron All-Sky Survey (2MASS), and/or
{\it Wide-Field Infrared Survey Explorer} ({\it WISE}) counterparts,
while a further 3, 10, 8, and 4 candidates are likely associated with
the Galactic Plane, nearby globular clusters, star-forming regions, or
local galaxies, respectively, even though they do not have clear,
single counterparts.
In total, only nine candidates show no sign of a counterpart to the
limits of these surveys, no association with extended objects, and/or a
location outside of the Galactic Plane (hereafter {\it criterion \#1}.

The benefit of using the CSC is that we can inspect individual data
products for each catalog source. To determine if the variability of
the CSC sources shows the same basic signatures as \hbox{CDF-S XT1}
(i.e., non-recurrent, $\lesssim$1\,ks rise, $\sim$$t^{-1.5}$ decline),
we visually inspected and classified the \hbox{X-ray} light curves of
all 184 sources into one of seven categories. Category 1 sources
appear to be recurrent or persistent transients, exhibiting single or
multiple flares, eclipses, and/or gradual variations on top of
otherwise quiescent rates. Recurrence/persistence was determined
directly from the Chandra data in some cases, or based on previously
known variability in the
SIMBAD\footnote{http://simbad.u-strasbg.fr/simbad/sim-fid} database
catalog. We consider category 1 sources to have distinctly different
variability behavior from \hbox{CDF-S XT1}.

Category 2--7 sources, on the other hand, were not considered to be
recurrent or persistent transients, with only marginal or no prior and
post detection of photons during the observations. We regard this as
a minimum requirement for similarity to \hbox{CDF-S XT1}, although we
caution that such a designation is strongly dependent on how
frequently a source is observed as well as the level to which an
assessment of variability is carried out. Cross-matching category 2--7
sources with all known X-ray archives and carefully investigating
variability across all possible instruments goes well beyond the scope
of this project, and would likely result in recategorization of at
least some fraction of these sources as Category 1 sources.

Beyond a basic estimate of recurrence or persistence, we further
divided the remaining Category 2--7 sources up based on their
variability properties. Category 2 sources exhibit $\gtrsim$4\,ks
rises and/or decay time-scales several times longer than \hbox{CDF-S
  XT1}. Category 3 sources exhibit $\sim$2--4 ks rise times and decay
timescales several times longer than CDF-S XT1 and/or flattening
either early or late in the transient decay. Category 4 sources
exhibit similar $\sim$2--4 ks rise times like category 3, but have
time series which terminate or are of insufficient statistical quality
to assess their decay rates properly. Given the strongly disparate
rise and decay timescales, category 2, 3 and 4 sources all seem
unlikely to be related to \hbox{CDF-S XT1}. 

Category 5 sources show X-ray light curve shapes consistent with that
of \hbox{CDF-S XT1} with peaks of $\lesssim$$10^{3}$\,s. As such, they
represent the most likely potential \hbox{CDF-S XT1} analogs. Category
6 sources are similar to category 5, but have significant low-level
activity in the $\sim$$10^{3}$--$10^{4.5}$\,s prior to their
``category 5'' light curve shapes. This precursor emission is not seen
in \hbox{CDF-S XT1} and may suggest that these objects have a
different physical origin and/or are recurrent or persistent. All
likely have Galactic origins. Therefore, they seem much less likely to
be potential \hbox{CDF-S XT1} analogs.  Finally, category 7 sources
exhibit weak short rise and fall behavior within $\sim$100--500\,s,
and are dominated by noise at late times. Thus, they could represent
fainter versions of \hbox{CDF-S XT1}. While two likely have Galactic
origins, one is associated with a faint SDSS+WISE galaxy with $z_{\rm
  ph}$$\sim$0.14 and another appears to be a strong radio source with
a WISE-only counterpart. Neither of the latter two identification
seems like a clear \hbox{CDF-S XT1} analog. In summary, after
assessing the X-ray light curves of all 184 CSC candidates (hereafter
{\it criterion \#2}), we find that category 5 and 7 objects may be
potential analogs, while all others categories show substantially
different variability behavior.

Finally, approximately 20--40\% of the 184 candidates have hardness
ratios or best-fitting spectral slopes consistent with $\Gamma<2.0$
(hereafter {\it criterion \#3}); this fraction also holds for the
candidates associated with categories 5 and 7.

\begin{figure*}
\vspace{0in}
\begin{center}
\hglue-0.2cm{\includegraphics[width=6.8in]{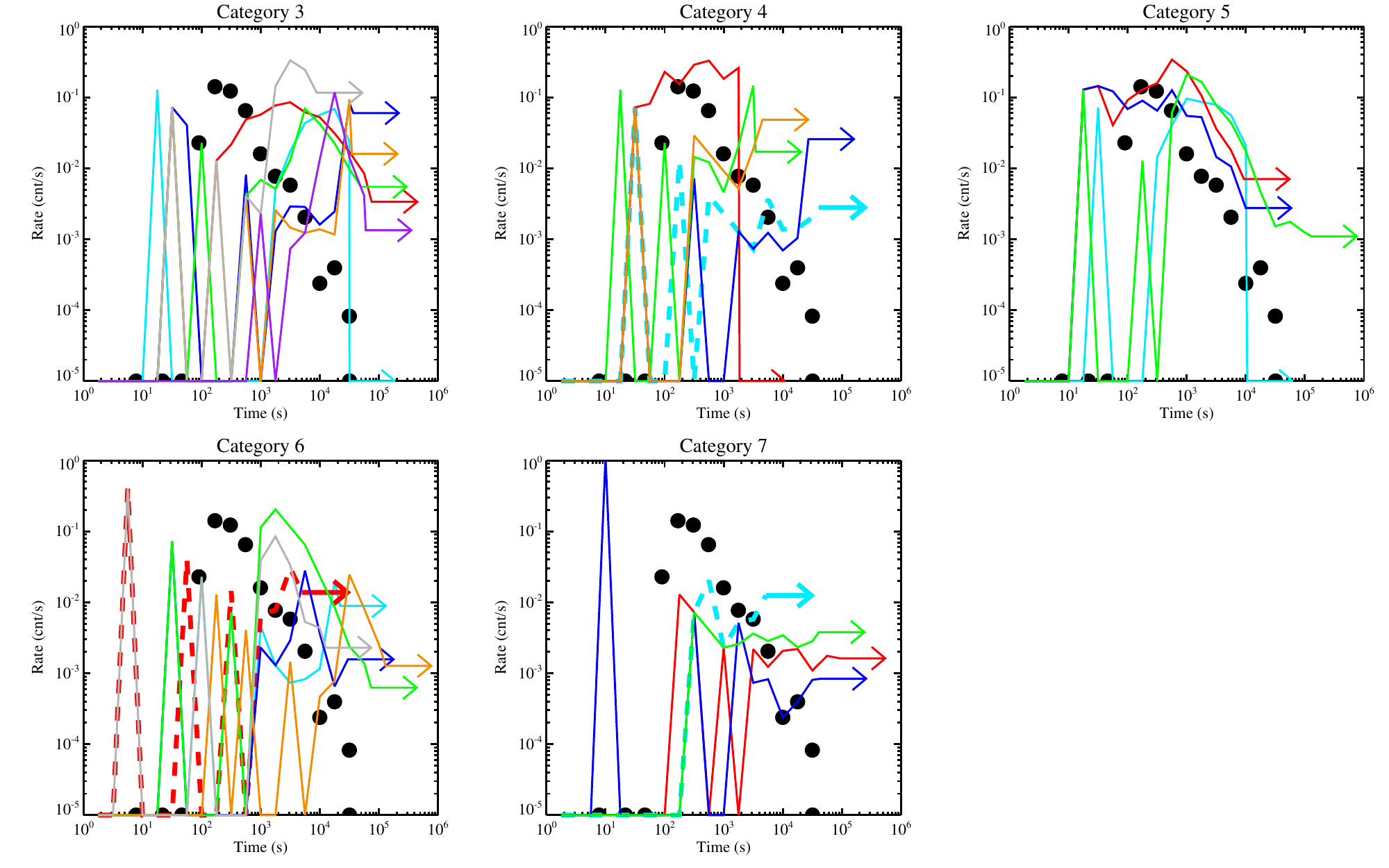}}
\end{center}
\vspace{0in}
\caption{\hbox{X-ray} light curves for the 26 CSC sources
    classified in categories 3--7, with each panel representing a
    given category. The CSC light curves are all binned on the same
    logarithmic scale and shown without error bars for clarity. The 23
    sources with and three sources without bright optical counterparts
    are denoted by thin solid and thick dashed lines,
    respectively. Based on counterpart identification, 21 sources
    likely have a Galactic origin.  The X-ray light curve of CDF-S XT1
    is shown by black circles in each panel, for comparison.  }
\vspace{-0.1in}
\label{fig:lc_all_csc}
\end{figure*}

A compilation of the results, broken down by category, is listed in
Table~\ref{tab:CSCsearch}, while the individual light curves of all
Category 3--7 sources are shown in Figure~\ref{fig:lc_all_csc}.
Factoring all three (imaging, timing and spectral) criteria together,
we find that there is not a single candidate that is comparable to
\hbox{CDF-S XT1}. For instance, among all 26 candidates in category
3--7, 21 sources appear to have a Galactic origin. Even amongst the
most likely candidates in categories 5--7 which show similar light
curves, roughly half have spectral slopes which are too soft and all
have an obvious Galactic or bright nearby galaxies origin. We thus
conclude that transients like \hbox{CDF-S XT1} appear to be rare or
alternatively hard to find based on relatively simple selection
criteria.}

We note that \citet{Glennie2015a} recently reported the detection of
two unusual high-amplitude, fast \hbox{X-ray} transients (FXRT) in the
{\it Chandra} archive, FXRT\,110103 and FXRT\,120830. However, both of
these exhibit strong \hbox{X-ray} flare behavior above constant
quiescent \hbox{X-ray} emission of
$\sim$10$^{-13}$--10$^{-12}$\,erg\,s$^{-1}$ cm$^{-2}$, which is
inconsistent with the \hbox{X-ray} behavior (strong quiescent limits)
from \hbox{CDF-S XT1}. FXRT\,120830 appears to be associated with a
bright flare from a nearby late M or early L dwarf star, and thus
appears to have a Galactic origin.  FXRT\,110103 lies at high Galactic
latitude ($b=32$\fdg7) and has no counterpart to $J>18.1$, $H>17.6$
and $K_{s}>16.3$. These limits are not too constraining and
FXRT\,110103 is tentatively associated by Glennie et al. with the
galaxy cluster ACO 3581 at 94.9 Mpc. \citet{Jonker2013a} also report
the discovery of FXRT\,000519, which has similar \hbox{X-ray} light
curve properties to FXRT\,110103 (including apparent quiescent
emission at a flux level of
$\sim$10$^{-13}$--10$^{-12}$\,erg\,s$^{-1}$ cm$^{-2}$) and is
associated with M86 at 16.2 Mpc.  Jonker et al. favor a tidal
disruption of a WD by an IMBH, but cannot rule out alternative
scenarios such as the accretion of an asteroid by a foreground NS or
an off-axis GRB. Given their associations with relatively nearby
  galaxies, FXRT\,110103 and FXRT\,000519 are factors of
  $10^{3}$--$10^{5}$ less luminous than \hbox{CDF-S XT1}, and it is
  not immediately obvious how they might be manifestations of the same
  phenomenon as \hbox{CDF-S XT1}. Hence for the moment, we do not
  consider them to be similar.

\begin{figure*}
\vspace{0in}
\begin{center}
\hglue-0.2cm{\includegraphics[height=3.2in, angle=0]{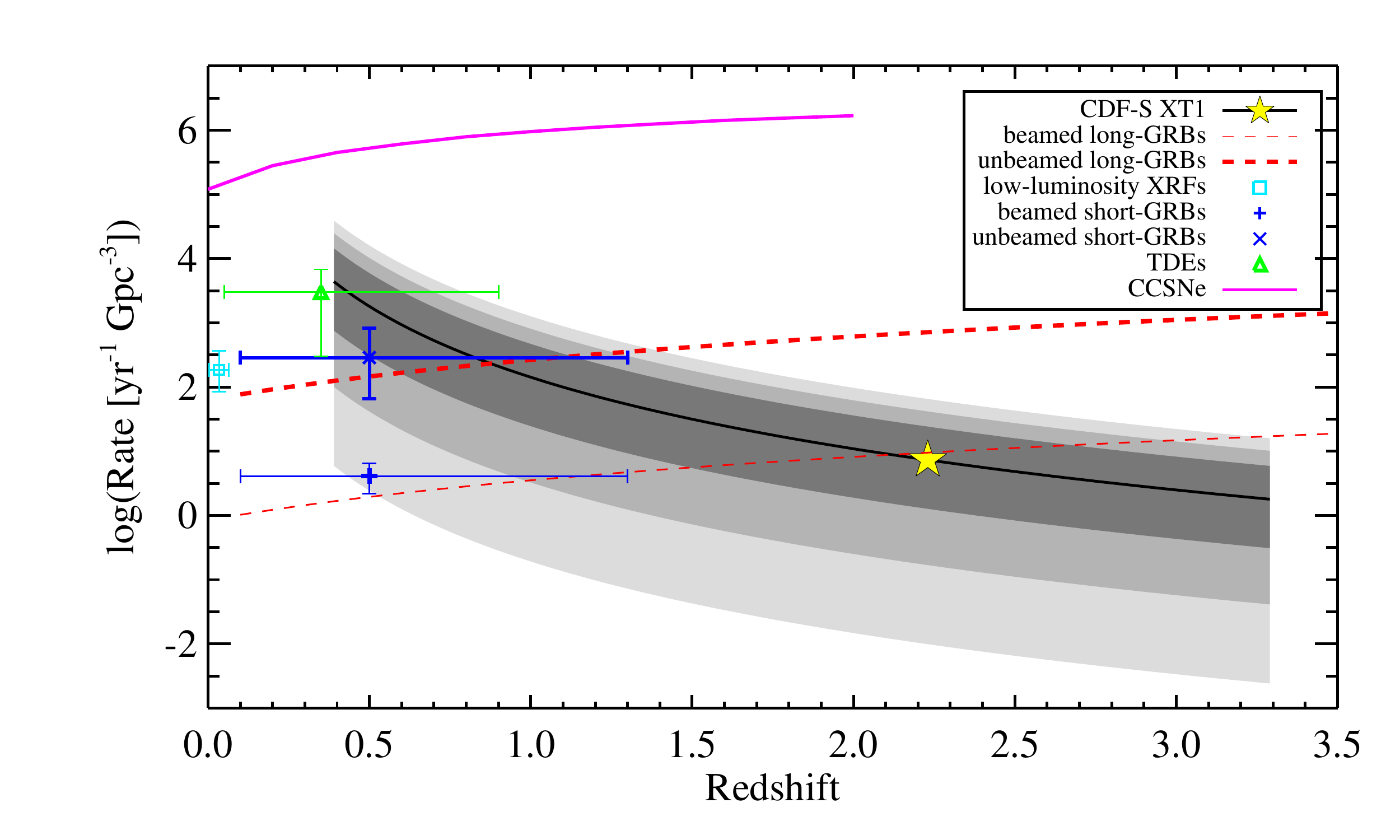}}
\end{center}
\vspace{0in}
\caption{Volumetric rates as a function of redshift for several
    known transient classes compared to \hbox{CDF-S XT1}. For
  \hbox{CDF-S XT1}, the black curve denotes the rate based on the
  volume enclosed at a given redshift, adopting a range spanning the
  95\% confidence limits of the photometric redshift of the associated
  host ($z_{\rm ph}$$=$0.39--3.21). The light, medium, and dark gray
  regions above and below the black curve denote the respective
  3$\sigma$, 2$\sigma$, and 1$\sigma$ rate ranges at a given
  redshift. Also shown are:
  a best-fit
  broken powerlaw model (thin dashed red curve) representing the
  intrinsic beamed lGRB rate from \citet{Wanderman2010a, Lien2014a};
  the intrinsic unbeamed (orphan) lGRB rate (thick dashed red curve) assuming a
  beaming correction of 75 \citep{Piran2004a, Guetta2005a};
  the observed rate of low-luminosity XRFs (cyan square);
  the intrinsic beamed sGRB rate (blue plus sign) from \citet{Wanderman2015a};
  the intrinsic unbeamed (orphan) sGRB rate (blue X sign) assuming a
  beaming correction of 70 \citep{Berger2014a};
  the intrinsic CCSNe rate \citep{Lien2009a};
  and the intrinsic TDE rate \citep{Stone2016a}.}
\label{fig:rates}
\end{figure*}

\subsection{Rate Estimation}\label{sec:rate_est}

Under the above criteria, we find no transients similar to \hbox{CDF-S
  XT1} in the entire CSC. If we relax some of the criteria, for
instance to allow for bright galaxy counterparts and a broader range
of light curve peaks, there are a few additional potential candidates
which could boost the rate by up to a factor of 3--4. However, until
there is a stronger understanding of the physics involved, we prefer
to remain conservative. With only one detected source, the rate of
such events is subject to large uncertainties. We adopt the above
characterization of the CSC as our baseline and estimate the coverage
of the CSC as follows.

For {\it{Chandra}}, there are two main configurations, ACIS-I and
ACIS-S, with as many as six 8\farcm5$\times$8\farcm5 detectors
operational for each; in recent years, {\it{Chandra}} has advocated
that users turn off at least two detectors. Alternatively, the HRC has
a field-of-view (FOV) of $\approx$30$'$$\times$30$'$. The point spread
function and effective area of all these detectors degrade
substantially beyond a few arcminutes from the aimpoint. For
reference, on-axis and with the current sensitivity, a source with
$F_{\rm 0.3-10\,keV}=4.2\times10^{-13}$~erg\,s$^{-1}$ cm$^{-2}$ and
the spectral slope of the transient will yield $\approx$100 counts for
ACIS-I, $\approx$150 counts for ACIS-S, and $\approx$64 counts for HRC
in 5.0\,ks, respectively. Given the degradation in sensitivity over
the lifetime of {\it{Chandra}}, early cycles would have detected
$\approx$1.5 times more photons. At 15$'$ off-axis, vignetting alone
will decrease the photon yields by $\approx$40\% for a source with
$\Gamma=1.5$, while the 90\% encircled energy PSF area will likewise
grow by a factor of $\approx$400, such that considerably more
background is included. The combination of these effects makes
reliable detection of a transient like \hbox{CDF-S XT1} difficult
beyond the primary four detectors I0-I3 for ACIS-I (289 arcmin$^{2}$),
the S2+S3+S4 detectors for ACIS-S (217 arcmin$^{2}$), or the central
100 arcmin$^{2}$ for HRC.  We will assume the shortest {\it{Chandra}}
exposure, $\approx$2\,ks, is sufficient to detect such a transient
(and thus all {\it{Chandra}} exposures are useful), although over such
short intervals it may be difficult to characterize the light curve
properly.

Summing up the total on-sky exposure time (livetime) examined in the
above CSC query up to 2010 August 10 with $|b|$$>$10$^{\circ}$, there
is 46.6\,Ms, 62.1\,Ms, and 3.7\,Ms for the ACIS-I, ACIS-S, or HRC
detectors, respectively.

Such exposures imply a total potential occurrence rate of up to
$4.2^{+9.7}_{-3.4}$ events deg$^{-2}$ yr$^{-1}$, adopting errors
following \citet{Gehrels1986a}. We convert this to a volumetric rate
(\,yr$^{-1}$ Gpc$^{-3}$) in Fig.~\ref{fig:rates}, assuming an
increasing volume as a function of redshift between the 95\%
confidence bounds of the photometric redshift of the associated
host. Following \citet{Sun2015a}, we quote event rates at the
  minimum peak luminosity to which they are probed. The {\it Chandra}
  observations within our CSC archive search should allow detection of
  ``\hbox{CDF-S XT1}''-like events to a peak flux limit of
  $\sim$(5--10)$\times$10$^{-13}$~erg\,s$^{-1}$ cm$^{-2}$, or a peak
  X-ray luminosity of $\sim$$10^{44}$--$10^{46}$~erg\,s$^{-1}$ over a
  redshift range of $z_{\rm ph}$$=$0.39--3.21.

Although we have already largely excluded an identification with most
known types of transients, it is still informative to compare the
above rate to the expected rates of other major transients, such as
sGRBs and lGRBs, SNe, and TDEs.

For lGRBs, we adopt a simple broken power law shape to model the
intrinsic rate of beamed lGRBs \citep{Wanderman2010a, Lien2014a},
assuming they roughly trace the shape of the cosmic star formation
rate as shown in Fig.~\ref{fig:rates}. At $z$$=$0, the rate is
anchored at a value of $\sim$0.84\,yr$^{-1}$ Gpc$^{-3}$ above a
  peak X-ray luminosity of $\sim$$10^{49}$\,erg\,s$^{-1}$ based on
simulations matched to the observed rate of lGRBs from {\it Swift}/BAT
\citep{Lien2014a}, with an uncertainty within a factor of $\sim$2.
Given that lGRBs are thought to be beamed and highly anisotropic
\citep{Harrison1999a, Levinson2002a}, it may be more appropriate to
compare with the rate of unbeamed, or ``orphan'', lGRB explosions,
which should be larger by the inverse of the beaming factor. For
simplicity, we adopt a beaming correction of $\approx$75$\pm$25 based
on large samples of lGRBs \citep{Piran2004a, Guetta2005a}, although we
caution that these corrections are often non-trivial, with low- and
high-luminosity GRBs likely to have different average half-opening
angles. For several well-studied lGRBs, for instance, the beaming
corrections were estimated to lie between 450--500 \citep{Frail2001a,
  vanPutten2003a}, implying a much higher orphan rate and at least a
factor of $\sim$6--7 uncertainty.

Given some of the potential similarities between \hbox{CDF-S XT1} and
the low-luminosity XRFs 060218/SN 2006aj (145 Mpc), 080109/SN 2008D
(27 Mpc), and 100316D/SN 2010bh (263 Mpc), we estimate their
cumulative observed rate assuming a volume of 300 Mpc, a 10\,yr {\it
  Swift/BAT} search window with 90\% efficiency, a 2 steradian {\it
  Swift/BAT} FOV, and a 10\% detection rate based on complex trigger
criteria \citep{Lien2014a}. This yields a rate of
$\sim$185$^{+181}_{-100}$\,yr$^{-1}$ Gpc$^{-3}$ above a peak
  X-ray luminosity of $\approx$$10^{46}$\,erg\,s$^{-1}$, with the
quoted errors being purely statistical, and likely severely
underestimating systematic uncertainties. Such rates and luminosity
limits are roughly consistent with the lGRB beaming factors mentioned
above.

For sGRBs, there are fewer robust identifications and
characterizations, leaving rate estimates substantially more
uncertain.  Based on available samples of $\sim$20 objects, the
estimated observed rate is 4.1$^{+2.1}_{-1.9}$\,yr$^{-1}$\,Gpc$^{-3}$
above a peak X-ray luminosity of
  $\sim$$10^{49}$\,erg\,s$^{-1}$ \citep{Wanderman2015a}. The
beaming corrections lie in the range $\sim$70$\pm$40
\citep{Berger2014a}, implying an intrinsic unbeamed or orphan sGRB
rate of $\sim$290$^{+530}_{-230}$\,yr$^{-1}$ Gpc$^{-3}$ between
$z$=0.1--1.3 above a peak X-ray luminosity of
  $\sim$$10^{47}$\,erg\,s$^{-1}$.

For CCSNe, the rate at $z$$=$0 is estimated to be
$\sim$10$^{5}$\,yr$^{-1}$ Gpc$^{-3}$ above a peak X-ray
  luminosity of $\sim$$10^{44}$\,erg\,s$^{-1}$ and is expected to
track the cosmic star formation rate as shown in Fig.~\ref{fig:rates}
\citep{Dahlen2004a, Lien2009a, Taylor2014a}. The uncertainties are
likely within a factor of $\sim$2 \citep{Horiuchi2011a}. Based on the
expected \hbox{X-ray} luminosities for SBOs, the local ($z$$<$0.5)
rates provide the most sensible comparison.

Finally, for TDEs, we note that the rates are still highly uncertain
due to limited number of detections and large uncertainties in the
underlying assumptions, such that estimates range between
$\sim$300--6800\,yr$^{-1}$ Gpc$^{-3}$ \citep{Stone2016a} above a
  peak X-ray luminosity of $\sim$$10^{42}$\,erg\,s$^{-1}$. The
rates of TDEs accompanied by relativistic jets should be significantly
smaller and with much higher X-ray luminosity limits
\citep{Bower2013a, vanVelzen2016a}.

From Fig.~\ref{fig:rates}, one can see that the estimated rates for
transients like \hbox{CDF-S XT1} appear similar to some other
transient populations.  A major caveat here is that the luminosity
limits for these various transient populations are quite
different. After matching luminosity limits following \citet{Sun2015a}
and references therein, we find that the \hbox{CDF-S XT1} rate is most
similar to the rates of unbeamed/orphan and/or low-luminosity lGRBs
and sGRBs between $z$$\sim$0.4--2.0. This provides further indirect
evidence that \hbox{CDF-S XT1} may be somehow related to the GRB
phenomenon. Although the rates of ``\hbox{CDF-S XT1}''-like events
remain highly uncertain, we note that they are still likely to be
substantially more common than extremely luminous, beamed TDEs out to
moderate redshift ($z$$\sim$1--2).

\section{Conclusions}\label{sec:conclude}

To summarize, during the acquisition of the final 3\,Ms of the
observations of the CDF-S 7\,Ms survey, we detected an exceptional
X-ray transient event, \hbox{CDF-S XT1}, at high-significance. The
{\hbox{X-ray} light curve of \hbox{CDF-S XT1} shows a fast rise
  [$\approx$100$(1+z)^{-1}$\,s] and a power-law decay time slope of
  $a$$=$$-1.53$$\pm$$0.27$, with little spectral variation and a peak
  flux of $F_{\rm 0.3-10\,keV, peak}$$=$$5.1$$\times$$10^{-12}$
  erg\,cm$^{-2}$\,s$^{-1}$.  The average spectrum can be modeled as an
  absorbed power law with a spectral slope of
  $\Gamma$$=$$1.43^{+0.26}_{-0.15}$ and an absorption limit of $N_{\rm
    H}$$<$$1.5$$\times$$10^{21}(1$$+$$z)^{2.5}$ cm$^{-2}$. The
  location of the event shows no prior or subsequent X-ray emission,
  allowing us to place 0.3--10\,keV quiescent and precursor limits
  that are factors of $10^{5}$ and $10^{3}$ times fainter,
  respectively.

CDF-S XT1 is robustly matched, within
$\approx$0\farcs13$\pm$0\farcs26, to a single optical counterpart,
which lies in the CANDELS region and thus benefits from deep {\it
  HST}, {\it Spitzer}, and ground-based imaging. The host is a
resolved $m_{\rm R}$=27.5 mag galaxy at $z_{\rm ph}$$=$2.23
(0.39--3.21 at 2$\sigma$ confidence). At this nominal redshift,
the host SED is consistent with that of a $M_{\rm R}$$\sim$$-18.7$
mag, $\log{M/M_{\odot}}$$\sim$$8.0$$\pm$$0.2$,
$1.15\pm0.04$\,$M_{\odot}$ yr$^{-1}$ dwarf galaxy.
The inferred observed 2--10\,keV peak luminosity of the event at this
redshift is $(6.8^{+0.7}_{-1.6})$$\times$$10^{46}$ erg s$^{-1}$.

The combination of the X-ray light curve properties, non-recurrence to
deep quiescent X-ray limits, robust faint quiescent optical
counterpart (or limits if somehow not associated) and lack of
associated multi-wavelength (optical/NIR, radio, $>$10\,keV) transient
emission to sensitive limits appear to exclude nearly all known types
of Galactic and extragalactic X-ray variables and transients.  A few
theoretical possibilities remain: an ``orphan'' X-ray afterglow from
an off-axis sGRB with weak optical emission; a low-luminosity GRB at
high redshift with no prompt emission below $\sim$20\,keV rest-frame;
or a strongly beamed TDE involving an intermediate-mass black hole and
a white dwarf with little variability. We stress that each scenario
likely requires considerable fine-tuning to comply with all of the
constraints. We encourage more efforts to explore and limit
  parameter space for \hbox{CDF-S XT1}. This situation bears
parallels with the discovery of fast radio bursts
\citep{Lorimer2007a}, which represented a completely new source class,
discovered by chance, with no clearcut physical explanation. 
  ``\hbox{CDF-S XT1}''-like events are indeed related to
  unbeamed/orphan sGRBs, then they will be relevant as sources of GW
  emission.

After failing to find any events identical to \hbox{CDF-S XT1} in the
{\it Chandra} Source Catalog (comprised of the first 11 cycles of {\it
  Chandra}), we estimate a rate of ``\hbox{CDF-S XT1}''-like events as
$<$4.2$^{+9.7}_{-3.4}$\,events~deg$^{-2}$\,yr$^{-1}$.  Although highly
uncertain due to the small number statistics and wide photometric
redshift range of its associated host, this potential rate
  appears crudely comparable at matched luminosity to that of
  unbeamed/orphan and low-luminosity lGRBs and sGRBs between
  $z$$\sim$0.4--2.0, lending additional weight to a possible link with
  this transient class.  Alternatively, the rate appears substantially
  higher than that expected for extremely luminous, beamed TDEs at
  moderate redshifts ($z$$\sim$1--2), although this class of TDEs
  likewise suffers from small number statistics at present. Regardless
  of whether these events belong to an untapped regime for a known
  transient class, or represent a new type of variable phenomena, the
  predicted rates imply that ``\hbox{CDF-S XT1}''-like events should
  be a relatively common physical phenomenon that we are just
  beginning to observe or understand.

Although beyond the scope of the current work, the peak 0.3--10\,keV
flux of \hbox{CDF-S XT1} is sufficiently bright to be detected by
several of the currently operating X-ray observatories. This could
lead to the discovery of further similar transients and would
certainly place more stringent limits on the rate estimates.  For
instance, incorporating the remainder of the {\it Chandra} archive
would increase coverage by $\sim$50\%, while searching through the
archives of {\it XMM-Newton} (FOV$\approx$0.25 deg$^{2}$,
$\approx$260\,Ms observed over 16 years based on the master
observation list catalog) and {\it Swift} (FOV$\approx$0.15 deg$^{2}$,
$\approx$250\,Ms over 11 years based on the master observation list
catalog) could increase the areal$+$temporal coverage by factors of
$\sim$7.6 and $\sim$4.9, respectively. Moreover, these observatories
could last another 5--15 years, pending funding extensions.
Alternatively, a few upcoming X-ray observatories may be able to make
significant further progress. The {\it eROSITA} mission
\citep{Merloni2012a} has a FOV of $\approx$0.833~deg$^{2}$ and
sensitivity sufficient to detect transients like \hbox{CDF-S XT1} in
each one of its eight passes over the sky during the nominal four year
mission. Thus, {\it eROSITA} should effectively provide an equivalent
coverage in sky area per time to the current {\it XMM-Newton}
archive. Each individual ``pass'' will provide an average exposure of
$\sim$320\,s, built up from a few short exposures every 4 hours, such
that rapid triggers can be performed. Given the relatively weaker hard
energy response of {\it eROSITA} compared to {\it XMM-Newton},
however, rapid X-ray follow-up within 1--2 hrs will likely be
necessary to properly characterize transients. Another prospect is the
wide-field telescope planned for the Chinese Academy of Sciences' {\it
  Einstein Probe} \citep{Yuan2015a}, which will have a FOV of
60$^{\circ}$$\times$60$^{\circ}$ and sensitivity of $\sim$10$^{-11}$
in 1000\,s. While this may be insufficient to detect \hbox{CDF-S XT1}
outright, it could potentially detect bright versions
($\gtrsim$10$\times$) if they exist. Taken together, we thus might
expect at least a handful of ``\hbox{CDF-S XT1}''-like sources to be
discovered in the next decade.

Looking further into the future,
next-generation observatories like ESA's {\it Athena}
\citep{Barrett2013a}, the proposed {\it X-ray Surveyor}
\citep{Weisskopf2015a} or any other wide-field X-ray observatory aim
to provide FOVs of order $\sim$0.6--1~deg$^{2}$ and substantially
better sensitivity than {\it Chandra} or {\it XMM-Newton}, so they may
detect several dozen transients like \hbox{CDF-S XT1} over their
lifetimes. However, these observatories will be in a much better
position to characterize the light curves in detail and probe factors
of at least 10 deeper to study fainter (and perhaps more abundant)
versions of \hbox{CDF-S XT1}. In all of the above cases, ``\hbox{CDF-S
  XT1}''-like events will strongly benefit from rapid multi-wavelength
follow-up to help constrain their physical nature.

\section*{Acknowledgements}

We acknowledge the staffs of ESO, Gemini, and {\it{HST}}, and in
particular Nancy Levenson, Blair Conn, Rodolfo Angeloni, Rene Rutten,
Christian Hummel, Linda Schmidtobreick, Claus Leitherer, Denise
Taylor, and Shantavia Sturgis for their help in promptly accepting our
DDT requests, preparing the observations and carrying them out. We
thank Belinda Wilkes and the {\it Chandra} staff for help
investigating {\it{Chandra}} instrumental effects. We would like to
thank David Palmer, Hans Krimm, Ersin {G{\"o}{\u g}{\"u}{\c s}, Yuki
  Kaneko, Alexander J. van der Horst, Shri Kulkarni, and Avishay
  Gal-Yam for their help and stimulating conversations, and the {\it
    Fermi} LAT team for providing LAT data products. We also thank the
  anonymous referee for several comments which help improve the
  manuscript.

We acknowledge support from: CONICYT-Chile grants Basal-CATA
PFB-06/2007 (FEB, ET, SS), FONDECYT Regular 1141218 (FEB) and 1160999
(ET), FONDECYT Postdoctorado 3140534 (SS), PCCI 130074 (FEB),
``EMBIGGEN'' Anillo ACT1101 (FEB, ET);
the Ministry of Economy, Development, and Tourism's Millennium
Science Initiative through grant IC120009, awarded to The Millennium
Institute of Astrophysics, MAS (FEB, FF, SS); 
Swiss National Science Foundation Grants PP00P2\_138979 and PP00P2\_166159 (KS);
National Natural Science Foundation of China grant 11673010 (BL); Ministry of Science and Technology of China grant 2016YFA0400702 (BL);
support from TLS Tautenburg, MPE Garching and DFG Kl/766 16-1 and 16-3 (DAK); 
Chandra \hbox{X-ray} Center grant GO4-15130A (BL, WNB);
NASA through Hubble Cycle 22 grant HST-GO-14043 (FEB) awarded by
the Space Telescope Science Institute, which is operated by the
Association of Universities for Research in Astronomy, Inc., for NASA,
under contract NAS 5-26555;
World Premier International Research Center Initiative (WPI
Initiative), MEXT, Japan (KM); 
Japan Society for the Promotion of
Science (JSPS) KAKENHI Grant 26800100 (KM) and JSPS Open
Partnership Bilateral Joint Research Projects (KM);
973 Programs 2015CB857005 (JXW) and 2015CB857004 (YQX); 
CAS Strategic Priority Research Program XDB09000000 (JXW, YQX); 
CAS Frontier Science Key Research Program QYZDJ-SSW-SLH006 (JXW, YQX); 
NSFC-11421303 (JXW, YQX); 
National Thousand Young Talents program (YQX); 
NSFC-11473026 (YQX); 
and Fundamental Research Funds for the Central Universities (YQX).

The scientific results reported in this article are based in part or
to a significant degree on observations made by the Chandra \hbox{X-ray}
Observatory, on observations obtained at the Gemini Observatory
acquired through the Gemini Observatory Archive and processed using
the Gemini IRAF package, which is operated by the Association
of Universities for Research in Astronomy, Inc., under a cooperative
agreement with the NSF on behalf of the Gemini partnership: the
National Science Foundation (United States), the National Research
Council (Canada), CONICYT (Chile), Ministerio de Ciencia,
Tecnolog\'{i}a e Innovaci\'{o}n Productiva (Argentina), and
Minist\'{e}rio da Ci\^{e}ncia, Tecnologia e Inova\c{c}\~{a}o (Brazil),
on observations made with ESO Telescopes at the La Silla Paranal
Observatory under programme ID 294.A-5005, and
on observations made with the NASA/ESA Hubble Space Telescope,
obtained from the Data Archive at the Space Telescope Science
Institute, which is operated by the Association of Universities for
Research in Astronomy, Inc., under NASA contract NAS 5-26555. These
observations are associated with program HST-GO-14043.
This work made use of data supplied by the UK Swift
Science Data Centre at the University of Leicester.
This work also made use of the Rainbow Cosmological Surveys Database,
which is operated by the Universidad Complutense de Madrid (UCM),
partnered with the University of California Observatories at Santa
Cruz (UCO/Lick,UCSC).
This research has made use of NASA's Astrophysics Data System
Bibliographic Services.




\bibliographystyle{mnras}







\bsp	
\label{lastpage}
\end{document}